\documentstyle{arPrep}
\begin{document}
\input psfig.sty
\jname{Annu. Rev. Astron. Astrophys.}
\jyear{2004}
\jvol{42}
\ARinfo{1056-8700/04/0610-00}


\newcommand\lya{Ly$\alpha$~}
\newcommand\ljeans{\lambda_{\rm J}}
\newcommand\ie{i.e.~}
\newcommand\kjeans{k_{\rm J}}
\newcommand\mjeans{M_{\rm J}}
\newcommand\beq{\begin{equation}}
\newcommand\eeq{\end{equation}}
\newcommand\Msun{$M_\odot$}
\newcommand\lyasource{{\dot N_\alpha}}
\newcommand\beqa{\begin{eqnarray}}
\newcommand\eeqa{\end{eqnarray}}
\newcommand\xb{{\bf x}}
\newcommand\rb{{\bf r}}
\newcommand\vb{{\bf v}}
\newcommand\ub{{\bf u}}
\newcommand\kb{{\bf k}}
\newcommand\Omm{{\Omega_m}}
\newcommand\Ommz{{\Omega_m^{\,z}}}
\newcommand\Omr{{\Omega_r}}
\newcommand\Omk{{\Omega_k}}
\newcommand\Oml{{\Omega_{\Lambda}}}
\newcommand\nb{\bar{n}}
\newcommand\etal{et al.\ }
\newcommand\Ng{N_\gamma}
\newcommand\zr{z_{\rm reion}}
\newcommand\HI{\rm H\,I}
\newcommand\cN{c_{\rm N}}
\newcommand\kB{k_{\rm B}}
\newcommand\Ni{N_{\rm ion}}
\newcommand\fg{f_{\rm gas}}
\newcommand\fe{f_{\rm eject}}
\newcommand\fin{f_{\rm int}}
\newcommand\fw{f_{\rm wind}}
\newcommand\JWST{{\it JWST}}
\newcommand\la{\mathrel{\hbox{\rlap{\hbox{\lower4pt\hbox{$\sim$}}}\hbox{$<$}}}}
\newcommand\ga{\mathrel{\hbox{\rlap{\hbox{\lower4pt\hbox{$\sim$}}}\hbox{$>$}}}}
\def\ionAR#1#2{#1$\;${\small\rm #2}\relax}
\newcommand\sun{\hbox{$\odot$}}
\newcommand\farcs{\hbox{$.\!\!^{\prime\prime}$}}
\newcommand\arcmin{\hbox{$^\prime$}}

\title{The First Stars}

\markboth{Bromm \& Larson}{First Stars}

\author{Volker Bromm
\affiliation{Department of Astronomy, Harvard University, 60 Garden St.,
Cambridge, Massachusetts 02138; email: vbromm@cfa.harvard.edu}
Richard B. Larson
\affiliation{Department of Astronomy, Yale University, Box 208101,
New Haven, Connecticut 06520-8101; email: larson@astro.yale.edu}}

\begin{keywords}
cosmology, first stars, intergalactic medium
\end{keywords}

\begin{abstract}
We review recent theoretical results on the formation of the first
stars in the universe, and emphasize related open
questions. In particular, we discuss the initial conditions for
Population~III star formation, as given by variants of the cold
dark matter cosmology. Numerical simulations have investigated the
collapse and the fragmentation of metal-free gas, showing that
the first stars were predominantly very massive. The exact determination
of the stellar masses, and the precise form of the primordial
initial mass function, is still hampered by our limited understanding
of the accretion physics and the protostellar feedback effects.
We address the importance of heavy elements in bringing
about the transition from an early star formation mode dominated by
massive stars, to the familiar mode dominated by low mass stars, at
later times. We show how complementary observations, both at high
redshifts and in our local cosmic neighborhood, can be utilized to probe
the first epoch of star formation.  
\end{abstract}

\maketitle

\section{INTRODUCTION}

The first stars to form out of un-enriched, pure H/He gas marked
the crucial transition from a homogeneous, simple universe to a
highly structured, complex one at the end of the cosmic dark ages.
Extending the familiar scheme of classifying stellar populations
in the local universe (Baade 1944) to the extreme case of zero
metallicity, the first stars constitute the so-called
Population~III (e.g., Bond 1981, Cayrel 1986, Carr 1987, 1994). 
The quest for Population~III stars has fascinated astronomers
for many decades, going back to the first tentative ideas of 
Schwarzschild \& Spitzer (1953). Recently, the subject has attracted
an increased interest, both from a theoretical and observational perspective.
New empirical probes of the high redshift universe have become available, 
and our ability to carry out sophisticated numerical simulations has
improved dramatically. In this review, we attempt to summarize the current
state of this rapidly evolving field.

The first generation
of stars had important effects on subsequent galaxy formation 
(see Carr et al. 1984 for an early overview).
On the one hand, Population~III
stars produced copious amounts of UV photons to reionize the universe
(e.g., Tumlinson \& Shull 2000,
Bromm et al. 2001b, Schaerer 2002, 2003, Tumlinson et al. 2003, 
Venkatesan \& Truran 2003, Venkatesan et al. 2003). Recently, the
{\it Wilkinson Microwave Anisotropy Probe (WMAP)} has observed the 
large-angle polarization anisotropy of the cosmic microwave
background (CMB), thus constraining the total ionizing photon
production from the first stars (e.g., Cen 2003a,b, Ciardi et al. 2003, 
Haiman \& Holder 2003, Holder et al. 2003, Kaplinghat et al. 2003,
Kogut et al. 2003, Sokasian et al. 2003a,b,
Wyithe \& Loeb 2003a,b). The large optical depth to Thomson scattering
measured by {\it WMAP} has been interpreted as a signature of a substantial early
activity of massive star formation at redshifts $z\ga 15$ (see Section 5.1).
Secondly, the supernova (SN) explosions that ended the lives
of the first stars were responsible for the initial enrichment 
of the intergalactic medium (IGM) with heavy elements (e.g., 
Ostriker \& Gnedin 1996, Gnedin \& Ostriker 1997, Ferrara et al. 2000,
Madau et al. 2001, Mori et al. 2002, Thacker et al. 2002,
Bromm et al. 2003, Furlanetto \& Loeb 2003,
Mackey et al. 2003, Scannapieco et al. 2003,
Wada \& Venkatesan 2003, Yoshida et al. 2004).
An intriguing possibility unique to zero-metallicity massive stars
is the complete disruption of the progenitor in a pair-instability
supernova (PISN), which is predicted to leave no remnant behind
(e.g., Barkat et al. 1967, Ober et al. 1983, Bond et al. 1984, Fryer et al. 2001,
Heger \& Woosley 2002, Heger et al. 2003). This peculiar explosion mode
could have played an important role in quickly seeding the IGM with
the first metals (see Section 4.2).

To place the study of the first stars into the appropriate
cosmological context, one has to ask (see Section 2):
{\it When and how did the cosmic dark ages end?}
The dark ages denote the cosmic time between
the emission of the CMB half a million years
after the big bang, when the CMB photons shifted into the infrared,
so that the universe would have
appeared completely dark to a human observer, and the formation of the
first sources of light (see Barkana \& Loeb 2001, Loeb \& Barkana 2001,
Miralda-Escud\'{e}~2003 for comprehensive reviews of this cosmic epoch).
Historically, the evocative -- and now widely used --
dark age analogy was first introduced 
into the literature briefly after the ascent of non-baryonic, cold
dark matter scenarios (e.g., Sargent 1986, Rees 1993).
In the context of popular cold dark matter (CDM) models of hierarchical
structure formation, the first stars are predicted to have formed in
dark matter (DM) halos of mass $\sim 10^{6}M_{\odot}$ that collapsed at redshifts
$z\simeq 20-30$ (e.g., Tegmark et al. 1997, Yoshida et al. 2003a). 
The first quasars, on the other hand
(e.g., Umemura et al. 1993, Loeb \& Rasio 1994, Eisenstein \& Loeb 1995),
are likely to have formed in more massive host systems, at
redshifts $z \ga 10$ (Haiman \& Loeb 2001, Bromm \& Loeb 2003a), and certainly
before $z\sim 6.4$, the redshift of the most distant quasar known
(Fan et al. 2003).
Here, we will not discuss quasar formation in depth, and we refer the reader
to the references above.

The most fundamental question about the first stars is how massive
they typically were (see Section 3).
Results from recent numerical simulations of the collapse and
fragmentation of primordial gas clouds suggest that the first stars were
predominantly very massive, with typical masses $M_{\ast}\ga 100
M_{\odot}$ (Bromm et al. 1999, 2002,
Nakamura \& Umemura 2001, Abel et al. 2000, 2002).
The simulations have taught us the important lesson that
regardless of the detailed initial conditions, the primordial gas
attains characteristic values of temperature and density. These in
turn correspond to a characteristic fragmentation scale, given
approximately by the Jeans mass, and are explained by the microphysics
of molecular hydrogen (H$_{2}$) cooling (see Section 3.2).
The minihalos which host the formation of the first stars
have virial temperatures of order $\sim 1000$~K, below the
threshold for cooling due to atomic hydrogen lines ($\sim 10^4$~K).
Therefore, cooling and fragmentation inside these minihalos
is possible only 
via rotational-vibrational transitions of H$_{2}$, 
which can be excited even at these low temperatures (see Section 2 for details).

Despite the
progress already made, many important questions remain unanswered.
A particularly important one, for example, is: {\it What
is the functional form of
the primordial initial mass function (IMF)?}  Having
constrained the characteristic mass scale, still leaves undetermined
the overall range of stellar masses and the power-law slope which is
likely to be a function of mass (e.g., Yoshii \& Saio 1986, Larson 1998,
Nakamura \& Umemura 2001, 2002, Omukai \& Yoshii 2003).
In addition, it is presently not known
whether binaries or, more generally, clusters of zero-metallicity
stars, can form (see Section 3.2). This question has important implications
for the star formation process. If Population~III star formation typically
resulted in a binary or multiple system, much of the progenitor cloud's
angular momentum could go into the orbital motion of the stars in such a system.
On the other hand, if an isolated formation mode were to predominate,
the classical angular momentum barrier could be much harder to breach
(see Larson 2003). 

How was the epoch of Population~III star formation
eventually terminated (see Section 4)?
Two categories of negative feedback effects are likely to be important,
the first being radiative and the second 
chemical in nature (e.g., Ciardi et al. 2000b, Schneider et al. 2002a,
Mackey et al. 2003). The radiative feedback consists in soft UV photons
produced by the first stars photodissociating the rather fragile
H$_{2}$ molecules in the surrounding gas, thus suppressing the corresponding
H$_{2}$ cooling 
(e.g., Haiman et al. 1997, 2000, Omukai \& Nishi 1999,
Ciardi et al. 2000a, Nishi \& Tashiro 2000,
Glover \& Brand 2001). Massive Population~III stars could then no longer
form.
There is some debate, however, whether the radiative feedback from the first
stars could not have had
an overall positive sign 
(e.g., Ferrara 1998, Machacek et al. 2001, 2003, Oh 2001, Ricotti et al. 2001,
Cen 2003b). The formation of H$_{2}$ molecules is catalyzed
by free electrons, and whichever process increased their presence 
would therefore also boost the abundance of H$_{2}$.
For example, an early background of
X-ray photons, either from accretion onto black holes (e.g., Madau et al. 2003),
or from the remnants of Population~III SNe, could ionize the gas and thus
increase the abundance of free electrons (e.g., Haiman et al. 2000,
Venkatesan et al. 2001,
Glover \& Brand 2003, Machacek et al. 2003).
This debate is ongoing, and 
it is too early for definitive conclusions (see Section 4.1 for details).

The second feedback effect that will be important in terminating the 
epoch of Population~III star formation is chemical in nature, or,
more precisely: it is due to the enrichment of the primordial gas with
the heavy elements dispersed by the first SNe (see Section 4.2).
Numerical simulations of the fragmentation process
have shown that lower mass stars can only form out of gas that was already
pre-enriched to a level in excess of the `critical metallicity', estimated to be
of order $10^{-4}-10^{-3}$ the solar value
(Bromm et al. 2001a; see also Schneider et al. 2002a).
Depending on the nature of the first SN explosions, and in particular
on how efficiently and widespread the mixing of the metal-enriched ejecta
proceeds, the cosmic star formation will at some point undergo a fundamental 
transition from an early high-mass (Population~III) dominated mode
to one dominated by lower mass stars (Population~II).

We conclude our review with a discussion of a few select observational probes
of the nature of the first stars (Section 5).
Predicting the properties of the first stars is important for the
design of upcoming instruments, such as the {\it James Webb Space
Telescope\footnote{See http:// ngst.gsfc.nasa.gov.} (JWST)}, or the
next generation of large ($>10$m) ground-based telescopes.  The hope
is that over the upcoming decade, it will become possible to confront
current theoretical predictions about the properties of the first
sources of light with direct observational data. The increasing volume
of new data on high redshift galaxies and quasars from existing ground- and
space-based telescopes, signals the emergence of this new frontier in
cosmology.

\section{COSMOLOGICAL CONTEXT}

The establishment of the current standard $\Lambda$CDM
model for cosmological structure formation (see Kirshner 2003,
Ostriker \& Steinhardt 2003 for recent reviews) has provided a
firm framework for the study of the first stars. Within variants 
of the CDM model, where larger structures 
are assembled hierarchically through successive mergers of smaller 
building-blocks, the first stars are predicted to form in DM
minihalos of typical mass $\sim 10^{6} M_{\odot}$ at redshifts
$z\sim 20 - 30$ (e.g., Couchman \& Rees 1986). The virial temperatures
in these low-mass halos, $T_{\rm vir}\propto M^{2/3}(1+z)$ (Barkana \& 
Loeb 2001), are below the threshold, $\sim 10^{4}$~K, for efficient 
cooling due to atomic hydrogen (e.g., Hutchings et al. 2002,
Oh \& Haiman 2002). It was realized
early on that cooling in the low-temperature
primordial gas had to rely on molecular hydrogen instead 
(Saslaw \& Zipoy 1967, Peebles \& Dicke 1968).

Since the thermodynamic behavior of the primordial gas thus
is primarily controlled by H$_{2}$ cooling, it is crucial to
understand the non-equilibrium chemistry of H$_{2}$ formation and destruction
(e.g., Lepp \& Shull 1984, Anninos \& Norman 1996, Abel et al. 1997,
Galli \& Palla 1998, Puy \& Signore 1999).
In the absence of dust grains to facilitate their formation (e.g., Hirashita
\& Ferrara 2002),
molecules have to form in the gas phase. The most important
formation channel turns out to be the sequence: H + $e^{-}\rightarrow$
H$^{-} + \gamma$, followed by H$^{-}+$ H $\rightarrow$ H$_{2}+ e^{-}$
(McDowell 1961). The free electrons act as catalysts, 
and are present as residue
from the epoch of recombination (Seager et al. 2000),
or result from collisional ionization in accretion shocks
during the hierarchical build-up of galaxies
(e.g., Mac Low \& Shull 1986, Shapiro \& Kang 1987).
The formation of hydrogen molecules thus ceases when the free
electrons have recombined. 
An alternative formation channel relies on the intermediary H$^{+}_{2}$
with free protons as catalysts (e.g., Haiman et al. 1996, Abel et al. 1997).
The H$^{-}$ channel, however, dominates in most circumstances (e.g., 
Tegmark et al. 1997).
Calculations of H$_{2}$ formation
in collapsing top-hat overdensities, idealizing the virialization
of dark matter halos in CDM cosmogonies, have found a simple
approximate relationship between the asymptotic H$_{2}$ abundance and
virial temperature in the overdensity (or halo): $f_{\rm H_{2}}
\propto T_{\rm vir}^{1.5}$ (Tegmark et al. 1997).

\begin{figure}[t]
\centerline{\psfig{figure=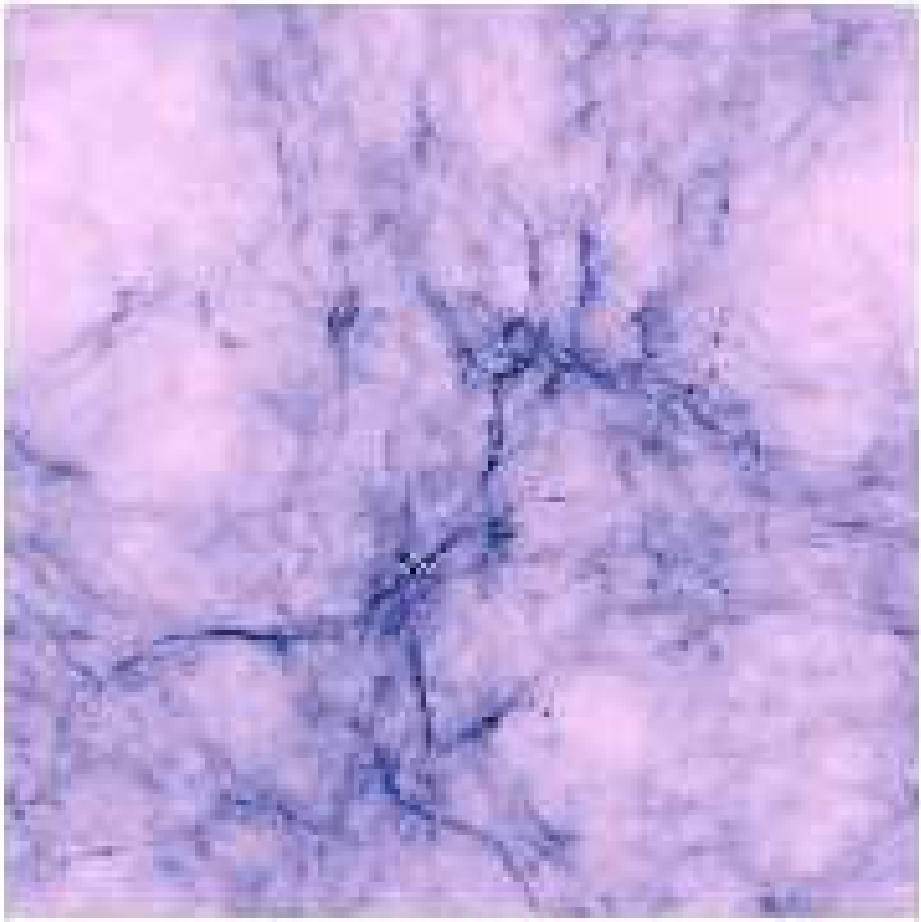,height=5in}}
\caption{Emergence of primordial star forming regions within
a standard $\Lambda$CDM cosmology. Shown is the projected gas density
at $z=17$ within a simulation box of physical size $\sim 50$~kpc.
The bright knots at the intersections of the filamentary network 
are the sites where the first stars form.
(From Yoshida et al.\ 2003a.)}
\label{fig1}
\end{figure}

Applying the familiar criterion (Rees \& Ostriker 1977, Silk 1977)
for the formation of galaxies
that the cooling timescale has to be shorter than the dynamical
timescale, $t_{\rm cool} < t_{\rm dyn}$, one can derive
the minimum halo mass at a given redshift inside of which the gas
is able to cool and eventually form stars (e.g.,
Tegmark et al. 1997, Santoro \& Thomas 2003).
The H$_{2}$ cooling
function has been quite uncertain, differing by an order of magnitude
over the relevant temperature regime (Hollenbach \& McKee 1979,
Lepp \& Shull 1984). Recent advances in the quantum-mechanical
computation of the collisional excitation process (H atoms colliding
with H$_{2}$ molecules) have provided a much more reliable
determination of the H$_{2}$ cooling function (see Galli \& Palla 1998, and
references therein).
Combining the CDM prescription for the assembly of virialized
DM halos with the H$_{2}$ driven thermal evolution of the primordial
gas, a minimum halo mass of $\sim 10^{6} M_{\odot}$
is required for collapse redshifts $z_{\rm vir}\sim 20-30$.
From detailed calculations, one finds that
the gas in such a `successful' halo has reached 
a molecule fraction in excess of
$f_{\rm H_{2}}\sim 10^{-4}$ (e.g., Haiman et al. 1996, 
Tegmark et al. 1997, Yoshida et al. 2003a).
These systems correspond to $3-4 \sigma$ peaks in the Gaussian
random field of primordial density fluctuations.
In principle, DM halos that are sufficiently massive to harbor
cold, dense gas clouds could form at higher redshifts, $z_{\rm vir}\ga 40$. Such systems, however,
would correspond to extremely rare, high-$\sigma$ peaks in the
Gaussian density field (e.g., Miralda-Escud\'{e} 2003).

To more realistically assess the formation of cold and dense star
forming clouds in the earliest stages of cosmological structure
formation, three-dimensional simulations of the combined
evolution of the DM and gas are required within a cosmological
set-up 
(Ostriker \& Gnedin 1996,
Gnedin \& Ostriker 1997, Abel et al. 1998).
These studies confirmed the important
role of H$_{2}$ cooling in low-mass halos at high $z$. Recently,
the problem of forming primordial gas clouds within a fully
cosmological context has been revisited with high numerical
resolution (Yoshida et al. 2003a,b,c,d). The resulting gas density
field is shown in Figure~1 for a standard $\Lambda$CDM cosmology at $z=17$.
The bright knots at the intersections of the filamentary network
are the star forming clouds, having individual masses (DM and gas)
of $\sim 10^{6} M_{\odot}$. As is expected from the statistics of
the high-$\sigma$ peaks (e.g., Kaiser 1984), the primordial clouds
are predominantly clustered, although there are a few cases of
more isolated ones.

\begin{figure}[t]
\centerline{\psfig{figure=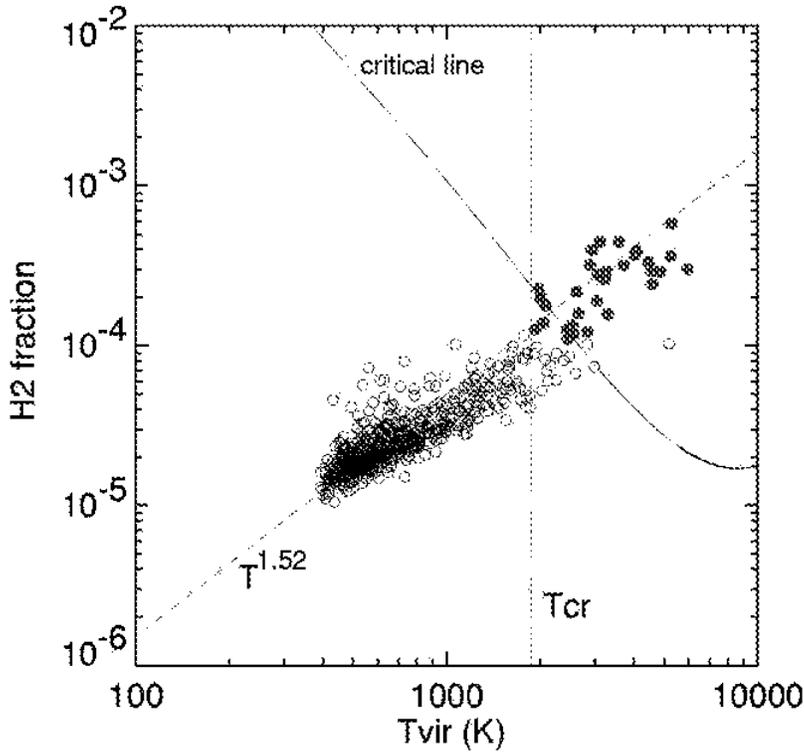,height=4in}}
\caption{Condition for successful formation of cold, dense gas
clouds. Mass-weighted mean H$_{2}$ fraction versus virial temperature.
({\it Open circles}:) DM halos that fail to host cold gas clouds.
({\it Filled circles}:) DM halos that succeed in harboring 
star-forming clouds. The critical line corresponds to the
condition $t_{\rm cool}\sim t_{\rm ff}$.
(From Yoshida et al.\ 2003a.)}
\label{fig2}
\end{figure}

Whether a given DM halo successfully hosts a cold, dense ($T\la 0.5 
T_{\rm vir}$,$n_{\rm H}\ga 5\times 10^{2}$cm$^{-3}$) gas cloud
can nicely be understood with the Rees-Ostriker criterion, as can be
seen in Figure~2. Yoshida et al. (2003a) derive a minimum collapse
mass of $M_{\rm crit}\simeq 7\times 10^{5}M_{\odot}$, with only
a weak dependence on collapse redshift (see also Haiman et al. 1996,
Fuller \& Couchman 2000, Machacek et al. 2001). 
The dynamical heating accompanying
the merging of DM halos has an important effect on the thermal and
chemical evolution of the gas. Clouds do not successfully cool
if they experience too rapid a growth in mass.

The primordial gas clouds that are found in the cosmological
simulations are the sites where the first stars form. It is,
therefore, important to learn what properties these clouds have
in terms of overall size, shape, and angular momentum contents.
The latter is often expressed by the familiar spin-parameter
$\lambda=L|E|^{1/2}/(G M^{5/2})$, where $L$, $E$, and $M$ are the
total angular momentum, energy, and mass, respectively. The spin
parameter is a measure of the degree of rotational support,
such that the ratio of centrifugal to gravitational acceleration
is given by $\sim \lambda^{2}$ at virialization. The spin values
measured in pure DM cosmological simulations
can be described by a lognormal distribution function with
a mean value of $\bar{\lambda}=0.04$, similar to what is found
for larger-scale systems (Jang-Condell \& Hernquist 2001).
Interestingly, the resulting angular momentum vectors for the
gaseous and DM components do generally not align in the simulations
of Yoshida et al. (2003a). The overall sizes of the Population~III
star forming clouds are close to the virial radius of the host
DM halo, with $R_{\rm vir}\sim 100$~pc, not too different
from the typical dimensions of present-day giant molecular clouds
(e.g., Larson 2003). Depending on the degree of spin, the clouds
have shapes with various degree of flattening (e.g., Bromm et al. 2002,
Yoshida et al. 2003a). Due to the importance of pressure forces, however,
overall cloud shapes tend to be rather spherical.
To fully elucidate the properties of the
star-forming primordial clouds, even higher resolution cosmological
simulations will be necessary.

The theoretical predictions for the formation sites of
the first stars sensitively depend on the exact nature of the
DM component and its fluctuation spectrum. Recently,
two models have been discussed that would much reduce the
fluctuation power on small mass scales. The first of these,
the warm dark matter (WDM) model (e.g., Bode et al. 2001), 
has been proposed to remedy
the well-known problems of standard $\Lambda$CDM on sub-galactic
scales (e.g., Flores \& Primack 1994, Moore et al. 1999).
These concern the predicted large abundance of substructure or,
equivalently, of satellite systems, and the high (cuspy) 
densities in the centers of galaxies. Both predictions are
in conflict with observations.
The second model, the `running' spectral index (RSI) model,
is suggested by the combined analysis of the {\it WMAP} data,
the 2dF galaxy redshift survey, and Lyman-$\alpha$ forest observations
(Spergel et al. 2003, Peiris et al. 2003). A series of recent 
studies have worked out the consequences of these reduced small-scale
power models on early star formation (Somerville et al. 2003,
Yoshida et al. 2003b,c). 
Within these models, the star formation rate at $z\ga 15$ is significantly
reduced compared to the standard $\Lambda$CDM case. This is 
due to the absence of low-mass halos and their associated gas clouds that are
cooled by molecular hydrogen.
We will revisit this issue when we discuss
the implications of the {\it WMAP} data for early star formation
(in Section 5.1). 

\section{FORMATION OF THE FIRST STARS}

\subsection{Star Formation Then and Now}

Currently, we do not have direct observational constraints on how
the first stars formed at
the end of the cosmic dark ages. It is, therefore, instructive to
briefly summarize what we have learned about star formation in the
present-day universe, where theoretical reasoning is guided by a
wealth of observational data (see Pudritz 2002, Ward-Thompson 2002, Larson 2003 
for recent reviews).

Population~I stars form out of cold, dense molecular gas that is
structured in a complex, highly inhomogeneous way. The molecular
clouds are supported against gravity by turbulent velocity fields and
pervaded on large scales by magnetic fields.  Stars tend to form in
clusters, ranging from a few hundred up to $\sim 10^{6}$ stars. It
appears likely that the clustered nature of star formation leads to
complicated dynamics and tidal interactions that transport angular
momentum, thus allowing the collapsing gas to overcome the classical
centrifugal barrier (Larson 2002).  The IMF of Population~I stars is
observed to have the approximate Salpeter form (e.g., Kroupa 2002)
\begin{equation}
\frac{{\rm d}N}{{\rm d log}M}\propto M^{-x} \mbox{\ ,}
\end{equation}
where
\begin{equation}
x\simeq \left\{
\begin{array}{rl}
1.35 & \mbox{for \ }M\ga 0.5 M_{\odot}\\
0.0 & \mbox{for \ } M\la 0.5 M_{\odot}\\
\end{array}
\right. \mbox{\ .}
\end{equation}
The IMF is not observationally well determined at the lowest masses,
but theory predicts that there should be a lower mass limit of about
$0.007 M_{\odot}$ set by opacity effects.
This theoretical limit reflects the minimum fragment mass, set when
the rate at which gravitational energy is released during the collapse
exceeds the rate at which the gas can cool (e.g., Low \& Lynden-Bell 1976,
Rees 1976).
The most important feature of the observed IMF is that $\sim 1
M_{\odot}$ is the characteristic mass scale of Population~I star formation,
in the sense that most of the mass goes into stars with masses close
to this value. Recent hydrodynamical simulations
of the collapse and fragmentation of turbulent  molecular cloud cores
(e.g., Padoan \& Nordlund 2002, Bate et al. 2003) illustrate the highly
dynamic and chaotic nature of the star formation process\footnote{See
http:// www.ukaff.ac.uk/starcluster for an animation.}.

The metal-rich chemistry, magnetohydrodynamics, and radiative
transfer involved in present-day star formation is complex, and we
still lack a comprehensive theoretical framework that predicts the IMF
from first principles. Star formation in the high redshift universe,
on the other hand, poses a theoretically more tractable problem due to
a number of simplifying features, such as: (i) the initial absence of
heavy elements and therefore of dust; (ii) the absence of
dynamically-significant magnetic fields in the pristine gas left over
from the big bang; and (iii) the absence of any effects from
previous episodes of star formation which would completely alter
the conditions for subsequent star formation.
The cooling of the primordial gas then depends only
on hydrogen in its atomic and molecular form.  Whereas in the
present-day interstellar medium, the initial state of the star forming
cloud is poorly constrained, the corresponding initial conditions for
primordial star formation are simple, given by the popular
$\Lambda$CDM model of cosmological structure formation. We now turn to
a discussion of this theoretically attractive and important problem.

\subsection{Clump Formation: The Characteristic Stellar Mass}

How did the first stars form? This subject has a long and venerable
history (e.g., Schwarzschild \& Spitzer 1953, Matsuda et al. 1969,
Yoneyama 1972, Hutchins 1976, Silk 1977, 1983, Yoshii \& Sabano 1979,
Carlberg 1981, Kashlinsky \& Rees 1983,
Palla et al. 1983, Yoshii \& Saio 1986). In this review, we focus
mainly on the more recent work since the renewed interest in high-redshift
star formation that began in the mid-1990s (e.g., Haiman et al. 1996,
Uehara et al. 1996, Haiman \& Loeb 1997, Tegmark et al. 1997, Larson 1998).

The complete answer to this question
would entail a theoretical prediction for the Population~III IMF,
which is rather challenging. A more tractable task is to estimate
the characteristic mass scale, $M_c$, of the first stars, and most
of the recent numerical work has focused on this simpler problem.
The characteristic mass is the mass below which the IMF flattens
or begins to decline (see Larson 1998 for examples of possible
analytic forms).
As mentioned above, this mass scale is observed to be $\sim 1
M_{\odot}$ in the present-day universe. 
Since the detailed shape of the primordial IMF is highly uncertain,
we focus here on constraining $M_c$, as this mass scale indicates
the typical outcome of the primordial star formation process, or
where most of the available mass ends up (e.g., Clarke \& Bromm 2003).

To fully explore the dynamical, thermal, and chemical properties of primordial
gas, three-dimensional numerical simulations are needed, although more
idealized investigations in two, one, or even zero (`one-zone models')
dimension are important in that they allow to probe a larger parameter space.
The proper initial conditions for primordial star formation are given
by the underlying model of cosmological structure formation (see Section 2).
It is, therefore, necessary to simulate both the DM and
gaseous (`baryonic') components. To date, two different numerical approaches
have been used to simulate the general three-dimensional fragmentation
problem in its cosmological context.

The first series of simulations used the adaptive mesh refinement (AMR)
technique (Abel et al. 2000, 2002; ABN henceforth). 
The AMR algorithm creates a nested
hierarchy of ever finer grids in fluid regions where high resolution
is needed (see, e.g., Balsara~2001 for a recent review).
This method allows to accurately simulate problems with high dynamic range,
i.e., large contrasts in density. Such a situation indeed occurs in 
primordial star formation where the proper initial conditions are
given by the large scale ($\la 10$~kpc) cosmological environment,
and where the collapsing gas has to be followed all the way to protostellar
scales ($\la 10^{-3}$~pc).
The second series of simulations used the smoothed particle hydrodynamics (SPH)
method (Bromm et al. 1999, 2002; BCL henceforth). 
This Lagrangian technique samples
the continuous fluid with a finite number of particles whose mass is
smoothed out according to a given distribution function (the SPH kernel),
and it has been applied to a wide range of problems in present-day
star formation
(see Monaghan~1992 for a standard review).

The SPH approach cannot compete with AMR in terms of dynamic range,
but it has the important advantage that it can easily accommodate
the creation of sink particles (e.g., Bate et al. 1995).
Quite generically, gravitational collapse proceeds very non-uniformly
such that the density in a few sub-regions becomes much larger than
the average. Numerically, this situation necessitates the adoption
of increasingly small timesteps: whereas the simulation follows
the evolution of the select density peak in detail, the overall system
is essentially frozen. To study the collective properties of star formation,
i.e., the formation of multiple high-density clumps that interact with each
other in a complex fashion (see Bate et al. 2003),
the SPH particles are merged into more
massive ones, provided they exceed a given density threshold.
Recently, the dynamical range of the standard SPH method has been
significantly improved by implementing a `particle splitting'
technique (Kitsionas \& Whitworth 2002, Bromm \& Loeb 2003a).
When the simulation reaches such high density in a certain region
that a sink particle would normally be created, a complementary strategy
is now adopted: Every SPH particle in the unrefined, high-density
region acts as a parent particle and spawns a given number of child particles,
and endows them with its properties. In effect, this is the SPH equivalent
of the grid-based AMR technique.

The most important difference between these two studies lies in the way
the initial conditions are implemented. The ABN simulations start at
$z=100$ with a realistic cosmological set-up, considering a periodic
volume of physical size $128/(1+z)$~kpc. The AMR technique allows ABN
to bridge the gap from cosmological to rotostellar scales. The BCL effort,
on the other hand, initializes the simulations, also at $z=100$, in a more
idealized way, as isolated, spherical overdensities that correspond to
high-$\sigma$ peaks in the Gaussian random field of cosmological density
fluctuations (see Katz 1991).

The systems considered by BCL comprise halos of total mass $10^{5}-10^{7}
M_{\odot}$ that collapse at $z_{\rm vir}\simeq 20 - 30$. The virialization
(or collapse) redshift, $z_{\rm vir}$, marks the approximate instant where
the DM component settles into virial equilibrium (leading to: 
$-E_{\rm grav}\sim 2 E_{\rm kin}$). The BCL initial configurations
are {\it not} simple `top-hat' (i.e., uniform density) clouds, however,
as the dynamically dominant DM component is endowed with density fluctuations
according to $P(k)\propto k^{-3}$, the asymptotic behavior of the
$\Lambda$CDM power spectrum on small mass scales (e.g., Yoshida et al. 2003d).
The BCL simulations, therefore, do reflect the basic bottom-up, hierarchical
merging of smaller DM halos into larger ones, and the accompanying shock
heating of the gas. Because of the isolated nature of their systems,
BCL had to treat the initial amount of angular momentum (or `spin')
as a free parameter, imparting both the DM and gaseous components
with solid-body rotation around the center of mass. In the ABN simulations,
tidal torques from neighboring halos generate the halo spin self-consistently.
In summary, the ABN investigation is the most realistic treatment of
the primordial star formation problem to date, whereas the more idealized
BCL simulations, implementing the same physical ingredients as ABN,
have explored a much wider range of initial conditions.

\begin{figure}[t]
\centerline{\psfig{figure=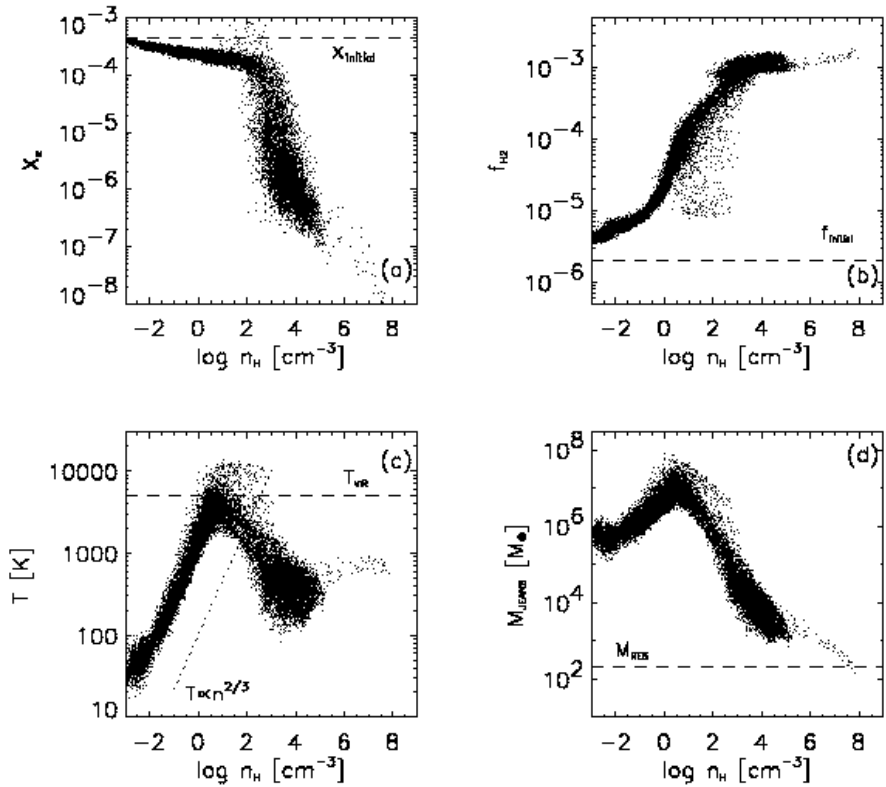,height=4.0in}}
\caption{
Properties of primordial gas.
{\bf (a)} Free electron abundance vs. hydrogen number density (in cm$^{-3}$).
{\bf (b)} Hydrogen molecule abundance vs. number density. 
{\bf (c)} Gas temperature vs. number density. At densities below $\sim 1$ cm$^
{-3}$, the gas temperature rises because of adiabatic compression until
it reaches the virial value of $T_{vir}\simeq 5000$ K.
At higher densities, cooling due to H$_{2}$
drives the temperature down again, until the gas settles into a quasi-
hydrostatic state at $T\sim 200$ K and $n\sim 10^{4}$ cm$^{-3}$.
Upon further compression due to the onset of the gravitational
instability, the temperature experiences a modest rise again.
{\bf (d)} Jeans mass (in $M_{\odot}$) vs. number density. The Jeans mass
reaches a value of $M_{J}\sim 10^{3}M_{\odot}$ for the quasi-hydrostatic
gas in the center of the DM potential well.
(From Bromm et al.\ 2002.)}
\label{fig3}
\end{figure}

In comparing the simulations of ABN and BCL, the most important aspect
is that both studies, employing very different methods, agree on the
existence of a preferred state for the primordial gas, corresponding
to characteristic values of temperature and density,
$T_c\sim 200$~K, and
$n_c\sim 10^{4}$ cm$^{-3}$, respectively. These characteristic scales
turn out to be rather robust in the sense that they are not very
sensitive to variations in the initial conditions.
In Figure~3 (panel c), this preferred state in the $T-n$ phase diagram
can clearly be discerned. This figure, from the simulations of BCL,
plots the respective gas properties for each individual SPH particle.
The diagram thus contains an additional dimension of information:
where evolutionary timescales are short, only few particles are
plotted, whereas they tend to accumulate where the overall evolution is slow.
Such a `loitering' state is reached at $T_c$ and $n_c$.

These characteristic scales can be understood by considering
the microphysics of H$_2$, the main coolant in metal-free, star forming
gas (Abel et al. 2002, Bromm et al. 2002). At temperatures $\la 1000$~K,
cooling is due to the collisional excitation and subsequent radiative 
decay
of rotational transitions. The two lowest-lying rotational energy levels
in H$_2$ have an energy spacing of $E/k_{\rm B}\simeq 512{\rm ~K}$. 
Collisions with particles (mostly H atoms) that populate the high-energy
tail of the Maxwell-Boltzmann velocity distribution can lead to somewhat
lower temperatures, but $H_2$ cooling cannot proceed to $T\la 100$~K.
This explains the characteristic temperature. The characteristic density,
in turn, is given by the critical density above which collisional
de-excitations, which do not cool the gas,  compete with 
radiative decays, which lead to cooling. This saturation
of the H$_2$ cooling marks the transition from NLTE rotational
level populations to thermal (LTE) ones. At densities below $n_c$, the
cooling rate is proportional to the density squared, whereas at higher
densities, the dependence is only linear.

Once the characteristic state is reached, the evolution towards higher
density is temporarily halted due to the now inefficient cooling, and the
gas undergoes a phase of quasi-hydrostatic, slow contraction.
To move away from this `loitering' regime, enough mass has to accumulate
to trigger a gravitational runaway collapse. This condition is simply
$M \ga M_J$, where the Jeans mass or, almost equivalently, the
Bonnor-Ebert mass can be written as (e.g., Clarke \& Bromm 2003)
\begin{equation}
M_J\simeq 700 M_{\odot}\left(\frac{T}{200{\rm ~K}}\right)^{3/2}
\left(\frac{n}{10^{4}{\rm cm}^{-3}}\right)^{-1/2} \mbox{\ .}
\end{equation}
Primordial star formation, where magnetic fields and turbulence are expected
to initially play no important dynamical role, may be the best case
for the application of the classical Jeans criterion which is describing
the balance between gravity and the opposing {\it thermal} pressure
(Jeans 1902).

A prestellar clump of mass $M\ga M_J$ is the immediate progenitor
of a single star or, in case of further subfragmentation, a binary
or small multiple system. In Galactic star-forming regions, like
$\rho$~Ophiuchi, such clumps with masses close to stellar values
have been observed as gravitationally bound clouds which lack the
emission from embedded stellar sources (e.g., Motte et al. 1998).
The high-density clumps are clearly not stars yet. To probe the
further fate of a clump, one first has to follow the collapse to
higher densities up to the formation of an optically thick hydrostatic
core in its center (see Section 3.3), and subsequently the accretion from
the diffuse envelope onto the central core (see Section 3.4). The parent
clump mass, however, already sets an upper limit for the final stellar
mass whose precise value is determined by the accretion process.

Although ABN and BCL agree on the magnitude of the characteristic
fragmentation scale, ABN have argued that only one star forms per halo,
while BCL have simulated cases where multiple clumps
form, such that the number of stars in a minihalo is $N_{\ast}\sim 1 - 5$.
Among the cases studied by BCL favoring $N_{\ast} > 1$, are high-spin runs that
lead to disk-like configurations which subsequently fragment into
a number of clumps, or more massive halos (with total mass $M \la 10^{7}
M_{\odot}$). Notice that the case simulated by ABN corresponds to
a gas mass of $\sim 3\times 10^{4} M_{\odot}$. Such a low-mass halo is
only marginally able to cool, and when BCL simulated a system with this mass,
they also found that only a single star forms inside the halo.
Further work is required to elucidate whether the multiple-clump formation
in BCL does also occur in simulations with realistic cosmological
initial conditions.

Whether indeed primordial clusters of metal-free stars are able to form
in the first galaxies is an important open question.
The answer will determine the proper interpretation of the primordial IMF.
If Population~III star formation were generally clustered, as is the case in the
present-day universe, the IMF would simply describe the actual distribution
of stellar masses in a given cluster. If, on the other hand, the first stars
formed predominantly in isolation, the IMF would
have the meaning of a probability distribution in a random process
that results in only one stellar mass per `draw'. The present-day IMF,
however, appears to be shaped largely by the chaotic interactions
between many accreting protostars in a common reservoir of gas (e.g., 
Bate et al. 2003). If there were no cluster to begin with, how would
the IMF be build up? Would there still be a self-similar extension
at masses $M > M_c$? Progress has to rely on improved simulations
in future work, and it is difficult to guess at the outcome.

A possible clustered nature of forming Population~III stars would also have
important consequences for the transport of angular momentum. In fact,
a crucial question is how the primordial gas can so efficiently shed
its angular momentum on the way to forming massive clumps.
In the case of a clustered formation process, angular momentum could
be efficiently transported outward by gravitational (tidal) torques.
Tidal torques can then transfer much of the angular momentum from
the gas around each forming clump to the orbital motion of the system,
similar to the case of present-day star formation in a clustered
environment (e.g., Larson 2002, Bate et al. 2003). In the isolated
formation process of ABN, however, such a mechanism is not available.
Alternatively, ABN have suggested the efficient transport of
angular momentum via hydrodynamic shocks during turbulent collapse.
Again, more work is required to convincingly sort out the angular momentum
transport issue. It may be possible that magnetic fields in an accretion
disk around the primordial protostar could be sufficiently amplified
by dynamo action from the suspected very weak primordial values (e.g.
Kulsrud et al. 1997).
The dynamo activity could occur as a consequence of turbulence that
is driven by gravitational instabilities in the disk (Tan \& Blackman 2003),
and magnetic stresses could then facilitate angular momentum transport
in the accretion disk.

As pointed out above, important contributions have also been made
by more idealized, two- to zero-dimensional, studies. 
These investigations typically ignore the dark matter component, thus
implicitly assuming that the gas has already dissipatively collapsed
to the point where the dark matter ceases to be dynamically dominant.
Much attention has been paid to the collapse of filamentary clouds
(Nakamura \& Umemura 1999, 2001, 2002).
Results from one- and two-dimensional simulations that otherwise 
include all the relevant processes for the primordial chemistry and
cooling have estimated fragmentation scales $M_c \sim 1 M_{\odot}$
and $M_c \sim 100 M_{\odot}$, depending on the (central) initial density.
When sufficiently high ($n \ga 10^5$cm$^{-3}$), the low-mass value is reached.
This bifurcation has led to the prediction of a bimodal IMF for the first
stars (Nakamura \& Umemura 2001). It is however, not obvious how
such a high initial density can be reached in a realistic situation
where the collapse starts from densities that are typically much smaller
than the bifurcation value.

Recently, the question of whether low mass metal-free stars can form
has received considerable interest in connection with the discovery
of extremely metal-poor (or more precisely: iron-poor) Galactic halo
stars (see Section 5.2). Is it possible to place a firm lower limit
on the mass of the first stars? Such a minimum mass may be given
by the so-called opacity limit for fragmentation
(e.g., Low \& Lynden-Bell 1976, Rees 1976).
The opacity limit, estimated to be $M_F\sim 0.007 M_{\odot}$ in present-day
star formation, is the mass of a cloud that is no longer
able to radiate away its gravitational binding energy in a free-fall time.
For the case of metal-free stars, one-dimensional, spherically
symmetric calculations have indicated a value which is rather similar
to the present-day opacity limit (Omukai \& Nishi 1998; see also Section 3.3).
Whereas Omukai \& Nishi (1998) have included all relevant sources of
opacity in metal-free gas, a somewhat more idealized study by Uehara et al.
(1996) has estimated that a cloud of $\sim 1 M_{\odot}$ may become opaque
to H$_{2}$-line radiation. 
It is currently not clear how relevant the latter value is in
a realistic collapse calculation.

We conclude this section with three other recent suggestions
of how to produce lower-mass stars out of zero-metallicity gas.
The first one is a modified version of the double-peaked (bimodal) IMF,
and relies on the presence of HD, deuterium hydride
(Nakamura \& Umemura 2002, Uehara \& Inutsuka 2000). In principle, HD is a much more efficient
coolant than H$_2$, as it possesses a permanent electric dipole moment
(with Einstein $A$ coefficients that are larger by a factor of $10^3$
compared to the quadrupole transition probability for H$_2$), and can cool
the gas to temperatures below $\sim 100$~K due to a smaller rotational
level spacing (e.g., Puy \& Signore 1996, Flower et al. 2000).
Bromm et al. (2002) included HD cooling in their simulations and found
that it does not change the thermal evolution of the gas. It may, however,
be possible to reach a regime where HD cooling becomes important when
the gas cools down from temperatures above 10,000~K, e.g., in photoionization
heated gas, or in shock-heated material in dwarf-galaxy size DM halos
(masses $\ga 10^8 M_{\odot}$).

The second idea postulates that the fragmentation scale (in effect the
Jeans mass) will be reduced in gas that is bathed in soft UV radiation,
below the high masses predicted for the first stars (Omukai \& Yoshii 2003). 
It is argued that the strong UV radiation in the vicinity of a massive
Population~III star will destroy H$_2$, thus preventing the gas to
reach the characteristic `loitering' state. The gas may then be able 
to collapse, albeit at higher temperatures, to increasingly large densities
(see also Omukai 2001).
The 3D-simulations of Bromm \& Loeb (2003a), treating a very
similar case, however, have failed to show the postulated fragmentation
to small masses.
The final idea proposed the formation of lower-mass stars
in zero-metal, shock compressed gas (Mackey et al. 2003).
The required compression could, e.g., be induced by the first energetic supernova
explosions. Assuming that the shock is radiative and isobaric (e.g., Shapiro
\& Kang 1987) one can derive the expression
\begin{equation}
M_J\simeq 10 M_{\odot}\left(\frac{n_0}{10^2{\rm ~cm}^{-3}}\right)^{-1/2}
\left(\frac{u_{\rm sh}}{200{\rm ~km s}^{-1}}\right)^{-1} \mbox{\ ,}
\end{equation}
where $n_0$ denotes the preshock density and $u_{\rm sh}$ the velocity of
the shock. Indeed, since a metal-free population of stars with a characteristic
mass intermediate between the `classical' Population~II and Population~III cases would be
quite distinct, Mackey et al. (2003) have suggested to name these
hypothetical stars `Population~II.5'. Numerical simulations of SN explosions
at high redshifts have so far failed to bear this prediction out
(Bromm et al. 2003), but it may be possible that Population~II.5
stars can only form in more massive systems (see Salvaterra et al. 2003).

We conclude this section by pointing out that all the realistic 3D
simulations of the first star formation problem to date have
consistently shown that the primordial gas fragments into
massive clumps. Conversely, none of the suggestions for making
low mass stars out of zero-metallicity gas has yet been realized in
a 3D simulation. The tentative conclusion therefore is that the
{\it typical} outcome of primordial star formation consists in massive 
clumps, and that low-mass clumps may need very special conditions to form.

\subsection{Protostellar collapse}

What is the further fate of the clumps discussed above?
In particular, one would like to test the notion that such a Jeans unstable
clump is the immediate progenitor of a single star which forms in its center.
That will evidently only be correct if the clump does not undergo further
subfragmentation upon collapsing to higher densities.
It has long been suspected that such subfragmentation could occur at
densities in excess of $\sim 10^{8}$cm$^{-3}$, at which point three-body
reactions become very efficient in converting the atomic gas (with only
a trace amount of H$_2$ molecules from the H$^-$ channel) into fully
molecular form: 3H $\rightarrow$ H$_2$ + H (Palla et al. 1983).
As the H$_2$ coolant is now suddenly more abundant by a factor of $\sim 10^{3}$,
the corresponding boost in cooling could trigger a thermal instability,
thus breaking up the clump into smaller pieces (Silk 1983).
Both ABN and BCL have included the three-body reactions in their chemical 
reaction networks, and have followed their simulations to higher densities
to test whether subfragmentation does occur or not (for BCL, this
calculation is described in Bromm 2000). Both groups report that no
further subfragmentation is seen. With hindsight, that may not be too
surprising. The reason being that any small density fluctuations which
are present earlier on, and which could serve as seeds for later fragmentation,
will have been erased by pressure forces during the slow, quasi-hydrostatic
`loitering' phase at $n \sim n_c$. 
In addition, inefficient cooling may also play a role in suppressing
high-density fragmentation. Despite the increase in the cooling rate 
throughout the fully molecular gas, this never
leads to a significant drop in temperature due to the countervailing effect
of compressional heating.

\begin{figure}[t]
\centerline{\psfig{figure=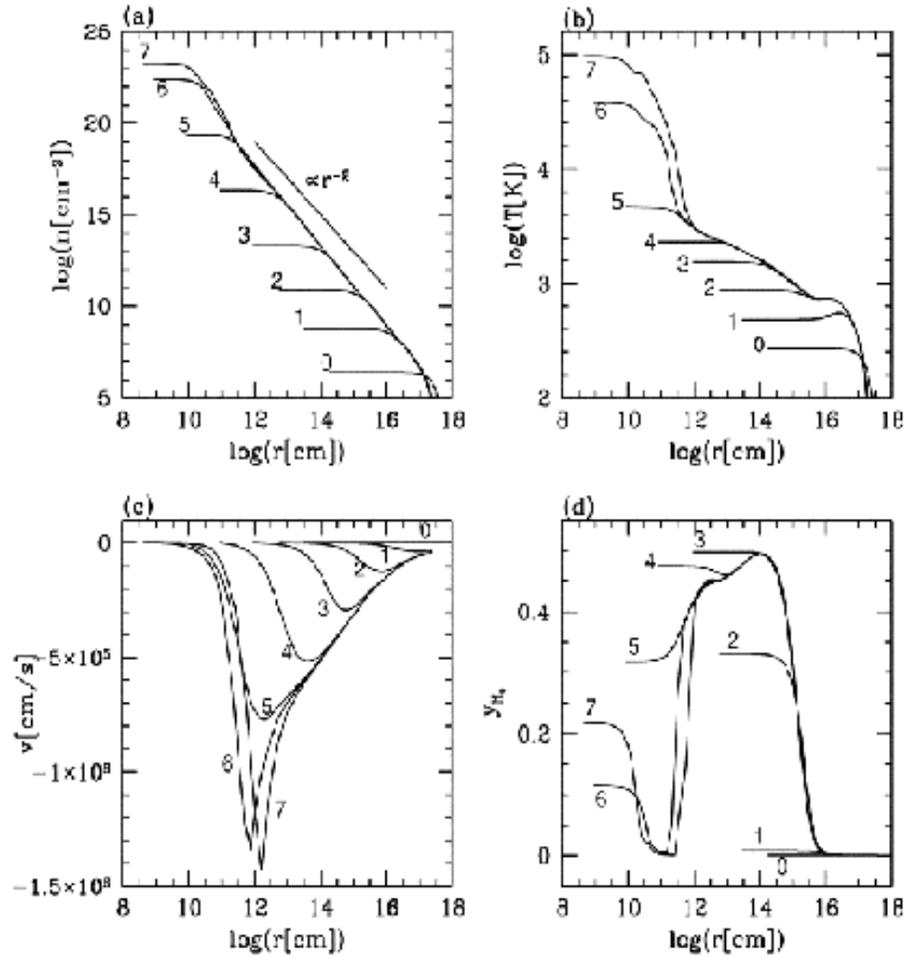,height=5.0in}}
\caption{
Collapse of a primordial protostar. Shown is the evolutionary sequence
as obtained in a 1D calculation, with time progressing from curves
labeled 1 to 7.
{\bf (a)} Number density vs. radial distance.
{\bf (b)} Gas temperature vs. distance. 
{\bf (c)} velocity vs. distance.
{\bf (d)} Hydrogen molecule abundance vs. distance.
The curves labeled 6 correspond to the situation briefly after the
formation of a central hydrostatic core.
(From Omukai \& Nishi 1998.)}
\label{fig4}
\end{figure}

Extending the analogous calculation for the collapse of a present-day
protostar (Larson 1969) to the primordial case, Omukai \& Nishi (1998)
have carried out one-dimensional hydrodynamical simulations in spherical
symmetry. They also consider the full set of chemical reactions, and
implement an algorithm to solve for the radiative transfer in the
H$_2$ lines, as well as in the continuum. The most important result is that
the mass of the initial hydrostatic core, formed in the center of the
collapsing cloud when the density is high enough ($n\sim 10^{22}$cm$^{-3}$)
for the gas to become optically thick to continuum radiation, is almost
the same as the (second) core in present-day star formation: $M_{\rm core}
\sim 5\times 10^{-3} M_{\odot}$. 
In Figure~4, the radial profiles
of density, temperature, velocity, and H$_2$ abundance are shown (reproduced
from Omukai \& Nishi 1998). The profiles of density and velocity
before the time of core formation (corresponding to the curves with label 6)
are well described by the Larson-Penston (LP) similarity solution. Once the
core has formed, the self-similarity is broken.
Similar results have been found by Ripamonti et al. (2002) who have 
in addition worked out the spectrum of the radiation that escapes
from the collapsing clump (mostly in the IR, both as continuum and line
photons). This type of approximately self-similar behavior seems to be
a very generic result of collapse with a simple
equation of state, even when rotation and magnetic fields are 
included (see Larson 2003). The apparent robustness of the LP solution
thus supports the results found for the Population~III case.

The small value of the initial core does not mean that this will be
the final mass of a Population~III star. This instead will be determined by
how efficient the accretion process will be in incorporating the clump mass
into the growing protostar. Before we discuss the accretion process in the
following section, we here briefly mention that fragments with masses
close to the  opacity limit could survive if the accretion
process were somehow prematurely curtailed. This mechanism has been found
in the simulation of a collapsing present-day, 
turbulent molecular cloud (Bate et al. 2003). A fraction of the accreting
cores is ejected, by slingshot interactions with neighboring ones, 
from the natal cloud before they could grow to stellar
masses, and thus end up as brown dwarfs (Bate et al. 2002). 
Again, such a mechanism would only work in a clustered environment, and as
discussed above, this might not be valid for primordial star formation.
An alternative suggestion to produce primordial brown dwarfs invokes
cooling due to HD which, as pointed out above,
could lead to lower temperatures and higher densities
(Uehara \& Inutsuka 2000).

\subsection{Accretion Physics}

\begin{figure}[t]
\centerline{\psfig{figure=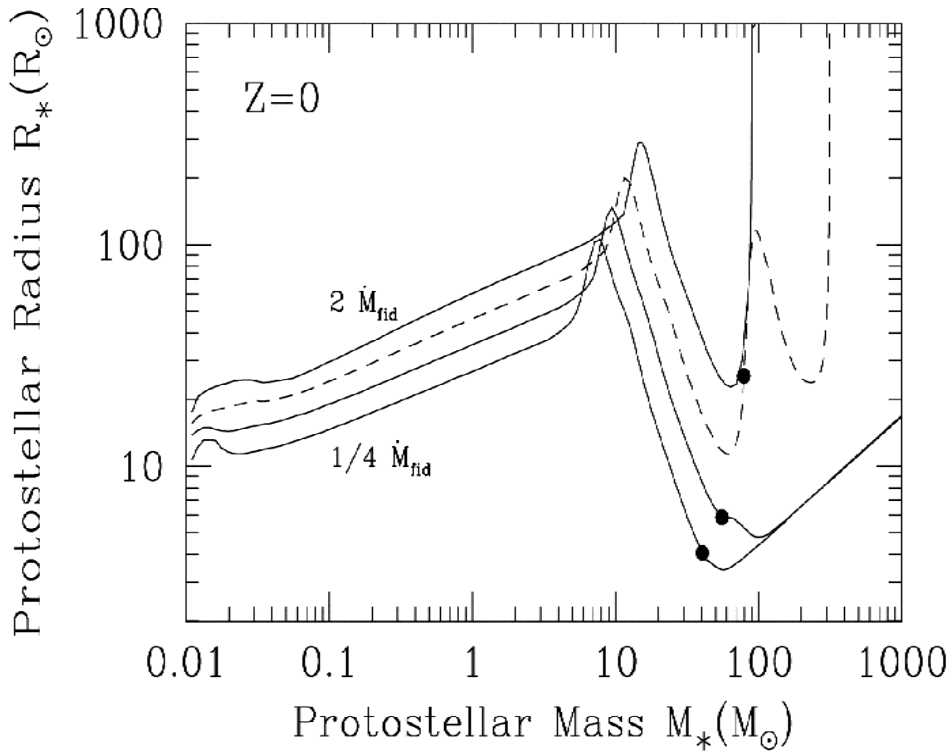,width=5.0in,height=4.0in}}
\caption{
Evolution of accreting metal-free protostar.
Shown is the radius-mass relation for different values of the
accretion rate (increasing from bottom to top). Accretion is
effectively shut off for the cases with $\dot{M}\ga \dot{M}_{\rm crit}$
because of the dramatic increase in radius.
(From Omukai \& Palla 2003.)}
\label{fig5}
\end{figure}

How massive were the first stars? Star formation typically
proceeds from the `inside-out', through the accretion of gas onto a
central hydrostatic core.  Whereas the initial mass of the hydrostatic
core is very similar for primordial and present-day star formation
(see above), the accretion process -- ultimately responsible for
setting the final stellar mass -- is expected to be rather different. On
dimensional grounds, the accretion rate is simply related to the sound
speed cubed over Newton's constant (or equivalently given by the ratio
of the Jeans mass and the free-fall time): $\dot{M}_{\rm acc}\sim
c_s^3/G \propto T^{3/2}$. A simple comparison of the temperatures in
present-day star forming regions ($T\sim 10$~K) with those in
primordial ones ($T\sim 200-300$~K) already indicates a difference in
the accretion rate of more than two orders of magnitude.

Extending earlier work by Stahler et al. (1986), Omukai \& Palla (2001,
2003) have investigated the accretion problem in considerable detail,
going beyond the simple dimensional argument given above.
Their computational technique approximates the time evolution
by considering a sequence of steady-state accretion flows onto a
growing hydrostatic core. Somewhat counterintuitively, these authors
identify a critical accretion rate, $\dot{M}_{\rm crit}\sim 4\times 10^{-3}
M_{\odot}$~yr$^{-1}$, such that for accretion rates higher than this,
the protostar cannot grow to masses much in excess of a few $10 M_{\odot}$.
For smaller rates, however, the accretion is predicted to
proceed all the way up to $\sim 600 M_{\odot}$, i.e., of order the
host clump.

The physical basis for the critical accretion rate is that
for ongoing accretion onto the core, the luminosity must not exceed
the Eddington value, $L_{\rm EDD}$. In the early stages of accretion,
before the onset of hydrogen burning, the luminosity is approximately 
given by $L_{\rm tot}\sim L_{\rm acc}\simeq G M_{\ast}\dot{M}_{\rm acc}/R_{\ast}$.
By demanding $L_{\rm acc}\simeq L_{\rm EDD}$ it follows
\begin{equation}
\dot{M}_{\rm crit}\simeq \frac{L_{\rm EDD}R_{\ast}}{G M_{\ast}}
\sim 5\times 10^{-3} M_{\odot}{\rm ~yr}^{-1} \mbox{\ ,}
\end{equation}
where $R_{\ast}\sim 5 R_{\odot}$, a typical value for a Population~III main-sequence
star (e.g., Bromm et al. 2001b). In Figure~5 (from Omukai \& Palla 2003),
the mass-radius relation is shown for various values of the
accretion rate. As can be seen, the dramatic swelling in radius
effectively shuts off  accretion at $M_{\ast}\la 100 M_{\odot}$, when
the accretion rate exceeds $M_{\rm crit}$.

Realistic accretion flows are expected to have a time-dependent rate,
and the outcome will thus depend on whether the accretion
rate will decline rapidly enough to avoid exceeding the Eddington
luminosity at some stage during the evolution.
The biggest caveat concerning the Omukai \& Palla results
seems to be the issue of geometry. A three-dimensional accretion
flow of gas with some residual degree of angular momentum
will deviate from spherical symmetry, and instead form a disk. It is
then conceivable that most of the photons can escape along the axes
whereas mass can flow in unimpeded through the accretion disk (see
Tan \& McKee 2003).

Recently, Bromm \& Loeb (2003b) have used the SPH particle splitting technique
to study the three-dimensional
accretion flow around a primordial protostar. 
Whereas BCL created sink particles already at $n\simeq 10^{8}$cm$^{-3}$, briefly
before three-body reactions would convert the gas into fully molecular form,
the gas is now allowed to reach
densities of $10^{12}$ cm$^{-3}$ before being incorporated into a
central sink particle. 
Bromm \& Loeb (2003b) studied how the molecular core grows in mass over the first
$\sim 10^{4}$~yr after its formation. The accretion rate
is initially very high, $\dot{M}_{\rm acc}\sim 0.1
M_{\odot}$~yr$^{-1}$, and subsequently declines according to a power
law, with a possible break at $\sim 5000$~yr. Expressed in terms of the
sound speed in Population~III prestellar cores, this initial rate
is:
$\dot{M}_{\rm acc}\sim 25
c_s^3/G $. Therefore, as initial accretion rates of a few tens times
$c_s^3/G $ are commonly encountered in simulations of present-day 
star formation, the Population~III case might again be an example of
a very generic type of behavior for collapse and accretion (see Larson 2003).

The mass of the
molecular core, taken as an estimator of the
protostellar mass, grows approximately as: $M_{\ast}\sim \int
\dot{M}_{\rm acc}{\rm d}t \propto t^{0.45}$. Rapidly, after only
$\sim 2000$~yr, the accretion rate drops below $\dot{M}_{\rm crit}$, at which
point the stellar mass is $M_{\ast}\sim 25 M_{\odot}$. Since in this case
the accretion rate becomes sub-critical early on,  
while the protostar still undergoes Kelvin-Helmholtz contraction (the 
descending portion of the curves in Figure~5), the dramatic expansion in radius,
which according to Omukai \& Palla (2003) would shut off further accretion, 
could be avoided.
A rough upper limit for
the final mass of the star is then: $M_{\ast}(t=3\times 10^{6}{\rm
yr})\sim 700 M_{\odot}$. This upper bound has been derived by
conservatively assuming that accretion cannot go on for longer than the
total lifetime of a very massive star (VMS), which is almost independent
of stellar mass (e.g., Bond et al. 1984).

{\it Can a Population~III star ever reach this asymptotic mass limit?}
The answer to this question is not yet known with any certainty, and
it depends on whether the accretion from the dust-free envelope is eventually
terminated by feedback from the star (e.g., Omukai \& Palla 2001, 2003,
Ripamonti et al. 2002, Omukai \& Inutsuka 2002, Tan \& McKee 2003).
The standard mechanism by which
accretion may be terminated in metal-rich gas, namely radiation pressure
on dust grains (Wolfire \& Cassinelli 1987), is evidently not effective for
gas with a primordial composition. Recently, it has been speculated
that accretion could instead be turned off through the formation of an
H~II region (Omukai \& Inutsuka 2002), or through the radiation pressure exerted
by trapped Ly$\alpha$ photons (Tan \& McKee 2003). The termination of the
accretion process defines the current unsolved frontier in studies of
Population~III star formation.

\subsection{The Second Generation of Stars: Critical Metallicity}

How and when did the transition take place from the
early formation of massive stars to that of low-mass stars at later
times?  
The very first stars, marking the cosmic Renaissance of structure
formation, formed under conditions that were much simpler than the
highly complex environment in present-day molecular clouds.
Subsequently, however, the situation rapidly became more complicated
again due to the feedback from the first stars on the IGM, both
due to the production of photons and of heavy elements (see Section 4).

In contrast to the formation mode of
massive stars (Population~III) at high redshifts, fragmentation is observed
to favor stars below a solar mass (Population~I and II) in the present-day
universe.  The transition between these fundamental modes is expected to be
mainly driven by the progressive enrichment of the cosmic gas with heavy
elements (see Section 4.2), which enables the gas to cool to lower temperatures.
The concept of a `critical metallicity', $Z_{\rm crit}$, has been
used to characterize the transition between
Population~III and Population~II formation modes, where $Z$ denotes the
mass fraction contributed by all heavy elements (Omukai 2000, 
Bromm et al. 2001a, Schneider et al. 2002a, 2003a, Mackey et al. 2003).
These studies have only constrained
this important parameter to within a few orders of magnitude, $Z_{\rm
crit}\sim 10^{-6}-10^{-3} Z_{\odot}$, under the implicit assumption of
solar relative abundances of metals.
This assumption is likely to be
violated by the metal yields of the first SNe at
high-redshifts, for which strong deviations from solar abundance ratios are
predicted (e.g., Heger \& Woosley 2002, Qian et al. 2002, Qian \& 
Wasserburg 2002, Umeda \& Nomoto 2002, 2003, Sneden \& Cowan 2003).
The cooling rate of the metals
depends on their ionization state, which is controlled by the ionizing
backgrounds (UV and X-ray photons or cosmic rays) and which is still
not well known (see Section 4.1).

Recently, Bromm \& Loeb (2003c)
have shown that the transition between the above star formation modes is
driven primarily by fine-structure line cooling of singly-ionized carbon or
neutral atomic oxygen.  Earlier estimates of $Z_{\rm crit}$ which
did not explicitly distinguish between different coolants are refined
by introducing
separate critical abundances for carbon and oxygen, [C/H]$_{\rm crit}$ and
[O/H]$_{\rm crit}$, respectively, where [A/H]= $\log_{10}(N_{\rm A}/N_{\rm
H})-\log_{10}(N_{\rm A}/N_{\rm H})_{\odot}$, and a subscript `$\odot$'
denotes solar values. Since C and O are also the most important coolants
throughout most of the cool atomic interstellar medium (ISM) in present-day
galaxies, it is not implausible that these species might be
responsible for the global shift in the star formation mode.
Indeed, under the temperature and density conditions that characterize
Population~III star formation, the most important coolants are O\,I and
C\,II whose fine-structure lines dominate over all other metal
transitions (Hollenbach \& McKee 1989). 
Cooling due to molecules becomes important only at
lower temperatures, and cooling due to dust grains only at higher
densities (e.g., Omukai 2000, Schneider et al. 2003a). The presence of
dust is likely to modify the equation of state at
these high densities in important ways (Schneider et al. 2003a), and it will
be interesting to explore its role in future collapse calculations.
The physical nature of the dust, however, that is produced by the first SNe
is currently still rather uncertain (e.g., Loeb \& Haiman 1997,
Todini \& Ferrara 2001, Nozawa et al. 2003, Schneider et al. 2003b).

A strong UV flux just below the
Lyman limit ($h\nu < 13.6$\,eV) is predicted to be emitted by
the same stars that are responsible for the reionization of the IGM (see
Section 4.1).
This soft UV radiation can penetrate the neutral IGM and ionize any trace amount
of neutral carbon due to its low first-ionization potential of
11.26\,eV. Since carbon is highly underabundant, the UV background can
ionize carbon throughout the universe well before hydrogen reionization.
Oxygen, on the other hand, will be predominantly neutral prior to
reionization since its ionization potential is 13.62\,eV.  Cooling is
mediated by excitations due to collisions of the respective metal with free
electrons or hydrogen atoms. At the low fractional abundances of electrons,
$x_{e}\la 10^{-4}$, expected in the neutral (rapidly recombining) gas at
$z\sim 15$, collisions with hydrogen atoms dominate. 
This renders the analysis independent
of the very uncertain hydrogen-ionizing backgrounds at high redshifts
(cosmic rays or soft X-rays).

To derive the critical C and O abundances, Bromm \& Loeb (2003c) start with
the characteristic state reached in primordial gas that cools only through
molecular hydrogen (Section 3.2).
For the gas to fragment further, 
additional cooling due to C\,II or O\,I is required. Fragmentation requires
that the radiative cooling rate be higher than the free-fall compressional
heating rate. The critical metal abundances can thus be found by equating the
two rates: $\Lambda_{\rm C\,II, O\,I}(n,T)\simeq 1.5 n k_{\rm B} T/t_{\rm ff}$,
where $k_{\rm B}$ is Boltzmann's constant.  
The critical C and O abundances are found by setting $n\sim n_c$ and $T \sim
T_c$, resulting in
[C/H]$_{\rm
crit}\simeq -3.5 \pm 0.1$ and [O/H]$_{\rm crit}\simeq -3.05 \pm 0.2$.
Strictly speaking, these threshold levels are the required abundances in
the gas phase. 

Even if sufficient C or O atoms are present to further cool the
gas, there will be a minimum attainable temperature
that is set by the interaction of the atoms with the thermal CMB: $T_{\rm
CMB}=2.7{\rm \,K}(1+z)$ (e.g., Larson 1998, Clarke \& Bromm 2003).
At $z\simeq 15$, this results in a characteristic
stellar mass of $M_{\ast}\sim 20 M_{\odot}(n_f/ 10^{4}{\rm
\,cm}^{-3})^{-1/2}$, where $n_f> 10^{4}{\rm \,cm}^{-3}$ is the density at
which opacity prevents further fragmentation (e.g., Rees 1976). 
It is possible that the transition from the high-mass to the low-mass
star formation mode was modulated by the CMB temperature and was therefore
gradual, involving intermediate-mass (`Population~II.5') stars
at intermediate redshifts (Mackey et al. 2003).
This transitional population could give rise to
the faint SNe that have been proposed to explain the observed abundance
patterns in metal-poor stars (Umeda \& Nomoto 2002, 2003).

When and how uniformly the transition in the cosmic star formation
mode did take place was governed by the detailed enrichment history
of the IGM. This in turn was determined by the hydrodynamical
transport and mixing of metals from the first SN explosions.
We will discuss the SN-driven dispersal of heavy elements in Section~4.2.

\section{FEEDBACK EFFECTS ON THE IGM}

\subsection{Radiative Feedback: The Suicidal Nature of the First Stars}

The radiation produced by Population~III star formation will influence
the subsequent thermal history of the IGM, and will modify the
properties of star forming gas. This radiative feedback may
occur in a variety of ways, depending on the energy range of the
stellar photons. 
As we have seen, molecular hydrogen is the main coolant in low temperature
metal-free gas, enabling the formation of the first stars at
$z\simeq 20-30$. We will therefore first
focus on the radiative effects that
are able to influence the H$_{2}$ chemistry.
This feedback
could be either negative, if due to soft UV photons, or
positive, if due to X-rays. 

We begin by discussing the negative feedback.
Molecular hydrogen is fragile and can readily be
destroyed by photons in the Lyman-Werner (LW) bands, within the energy
range $11.2-13.6$~eV, via the two-step Solomon process (Stecher \&
Williams 1967)
\begin{equation}
{\rm H}_{2} + \gamma \rightarrow {\rm H}_{2}^{\ast} \rightarrow 2 {\rm H} {\rm .}
\end{equation}
The intermediate stage involves an excited electronic 
state, ${\rm H}_{2}^{\ast}$, from which a fraction of the subsequent
decays end in the vibrational continuum of the ground state, resulting
in the dissociation of the molecule (e.g., Glover \& Brand 2001). 
The question arises whether H$_{2}$ cooling can be
suppressed in halos that virialize after the first stars have
formed and built up a background of LW photons (e.g., Haiman et al. 1997).
These photons can easily
penetrate a still neutral IGM prior to reionization (having energies
just below the Lyman limit). These second-generation
halos are typically more massive compared to the $\sim 10^{5} -
10^{6}M_{\odot}$ systems that host the formation of the first stars at
$z\ga 20$. The gas might then be able to self-shield against the
photo-dissociating LW background (e.g., Glover \& Brand 2001, Kitayama et 
al. 2001, Machacek et al. 2001). 
The effect of self-shielding is often taken into account, both in simulations 
and in analytical work,
by writing the
H$_{2}$ photodissociation rate as $k_{\rm diss}\propto J_{21}f_{\rm
shield}$, with a proportionality constant that depends on the spectrum of
the background radiation.
Here, $J_{21}$ measures the flux at the Lyman limit in the customary units of
$10^{-21}$~ergs s$^{-1}$ cm$^{-2}$ Hz$^{-1}$ sr$^{-1}$.
For the shielding factor, 
the following approximate expression is usually used:
$f_{\rm shield}\simeq \min[1,(N_{\rm H_{2}}/10^{14}{\rm
cm}^{-2})^{-0.75}]$ (Draine \& Bertoldi 1996). This formula is
accurate only for a static medium, and it will grossly overestimate
the H$_{2}$ line opacity in the presence of large-scale velocity
gradients. In cases with strong bulk
flows,
the effect of self-shielding is conservatively overestimated.

The overall normalization of the LW background, and its
evolution with redshift,
is still rather uncertain.
Close to the epoch of reionization,
however, a significant flux in the LW bands ($h\nu < 13.6$~eV) is
expected
(e.g., Bromm \& Loeb 2003a).
The flux in the LW bands just below the Lyman limit, $J_{21}^{-}$,
could be much larger than that just above, $J_{21}^{+}$. Assuming that only a
fraction, $f_{\rm esc}$, of the ionizing photons can escape from the
star forming halos, one has: $J^{-}_{21}\simeq J^{+}_{21}/f_{\rm
esc}$. This could be quite large, since $f_{\rm esc}$ is
expected to be low at high redshifts (Wood \& Loeb 2000, and
references therein). Alternatively, a strong background of photo-dissociating
photons could also arise internally
from star formation in an early dwarf galaxy
itself (e.g., Omukai \& Nishi 1999, Oh \& Haiman 2002). 

The strong UV background flux estimated above makes it possible for
H$_{2}$ formation to be effectively suppressed close to the redshift
of reionization.  In the absence of molecular hydrogen, however,
cooling can still proceed via atomic transitions in halos of mass (see
Barkana \& Loeb 2001)
\begin{equation}
M\ga
 10^{8} M_{\odot} \left(\frac{1+z}{10}\right)^{-3/2}
\mbox{\, .}
\end{equation}
The virial temperature, $T_{\rm vir}\sim 10^{4}$K, in these more
massive halos allows for the very efficient cooling of the gas via
lines of atomic hydrogen (e.g., Madau et al. 2001, Bromm \& Clarke 2002).  
In cases where the gas
temperature is close to the virial temperature, the gas cloud as a
whole can undergo collapse but it will not be able to fragment until
high enough densities are reached so that the Jeans mass has declined
sufficiently (Oh \& Haiman 2002, Bromm \& Loeb 2003a).
Recent numerical simulations of the collapse of a dwarf galaxy at
$z\simeq 10$, massive enough for atomic H cooling to
operate, but where cooling due to H$_2$ is effectively suppressed by
a strong LW background, have shown that very massive ($\sim 10^6 M_{\odot}$),
compact (radii $\la 1$~pc) gas clouds form in the center of the DM
potential well (Bromm \& Loeb 2003a). These clouds may be the immediate
progenitors of supermassive black holes that could seed high-redshift quasar
activity.

In contrast to this negative feedback on H$_{2}$, opposing positive
feedbacks have been proposed. As discussed in Section~2, H$_2$ formation
is primarily facilitated by free electrons. Any process which temporarily
enhances their abundance, therefore, will also tend to increase
the H$_2$ fraction. Various mechanisms have been suggested. First, 
X-ray photons from SN remnants (e.g., Oh 2001, Cen 2003a) or the
accretion onto black holes could ionize hydrogen in dense regions, enabling
the reformation of H$_2$ (Glover \& Brand 2003, Machacek et al. 2003).
Second, the free electrons in relic H~II regions (Oh \& Haiman 2003),
or inside the outer boundary of a still active one (Ricotti et al.
2001, 2002a, b), may accomplish the same. Finally, the collisional
ionization in shock waves, and the subsequent non-equilibrium 
cooling and molecule reformation has been discussed (e.g., 
Mac Low \& Shull 1986, Shapiro \& Kang 1987, Ferrara 1998, Bromm et al. 2003).
The last mechanism evidently is not strictly radiative in nature, but
it seems appropriate to include it here.
In summary, conclusions on the overall sign of the radiative feedback
on the H$_{2}$ chemistry
are still very uncertain. To reach more robust results, sophisticated
cosmological simulations are required, treating the build-up of 
the radiation backgrounds, and their influence on the overall star formation
rate in a self-consistent manner. Such challenging simulations are 
currently still beyond our capabilities.

A qualitatively different feedback
is the photoheating due to radiation with energies high enough
to ionize hydrogen atoms 
(see Barkana \& Loeb 2001 for a comprehensive discussion).
Since photoionization-heating raises the
gas temperature to $\ga 10^{4}$~K, this gas is then evaporated out
of the small potential wells at high redshifts (with virial temperatures 
$T_{\rm vir} < 10^{4}$~K), and star formation is thus temporarily
shut off in these halos. This photoevaporation feedback could be
due to sources in the halo itself (e.g., Bromm et al. 2003, 
Oh \& Haiman 2003, Whalen et al. 2003), or to an external ionizing background
(e.g., Barkana \& Loeb 1999, Susa \& Umemura 2000, Kitayama et al. 2001, 
Dijkstra et al. 2003, 
Shapiro et al. 2003, Tassis et al. 2003). The external feedback 
is likely to become important only at redshifts close to the complete
reionization of the universe at $z\sim 6$, whereas the internal one
could operate already at much higher redshifts.

\subsection{Chemical Feedback: First Supernovae and Early Metal Enrichment}

The crucial transition from a smooth, homogeneous universe to an increasingly
complex and structured one is marked by the death of the first stars.
The SN explosions at the end of the initial burst of Population~III
star formation disperse the first elements into the hitherto pristine
IGM.
How did this initial metal enrichment proceed?
To account for the widespread presence of metals in the Ly$\alpha$
forest at $z\sim 4 - 5$ (e.g., Songaila 2001, Schaye et al. 2003),
star formation in low-mass
systems at $z\ga 10$ has been proposed as a likely source 
(e.g., Madau et al. 2001, Mori et al. 2002, Wada \& Venkatesan 2003).
The metals produced in these low-mass
halos can more easily escape from their shallow potential wells
than those at lower redshift (e.g., Aguirre et al. 2001a,b).
In addition,
the enriched gas has to travel much shorter distances between neighboring
halos at these early times, and it might therefore have been easier to
establish a uniform metal distribution in the IGM.

The important question arises: {\it How did the
first stars die?}
The answer to this question sensitively depends on the precise mass of
a Population~III star. In particular, if the star has a mass in the narrow
interval $140 \la M_{\ast} \la 260 M_{\sun}$, it will explode as a
pair-instability supernova (PISN), leading to the complete disruption
of the progenitor (e.g., Fryer et al. 2001, Heger et al. 2003).
Population~III stars with masses below or above the PISN range are predicted
to form black holes. This latter fate is not accompanied by a significant
dispersal of heavy elements into the IGM, since
most of the newly synthesized metals will be locked up in the black
hole. The PISN, however, will contribute {\it all} its heavy element 
production to the surrounding gas. 

As we have discussed in Section~3.5,
the IGM metal enrichment history is crucial for understanding
the detailed transition in the star formation mode from high-mass to
low-mass dominated. If this enrichment is very uniform, the transition
would occur rather suddenly; if, on the other hand, enrichment would
be very patchy, with islands of high-metallicity gas embedded in mostly
still pristine material, then the transition would occur in a 
non-synchronized way, probably occupying a considerable time interval
(e.g., Scannapieco et al. 2003).

\begin{figure} [t]
\centerline{\psfig{file=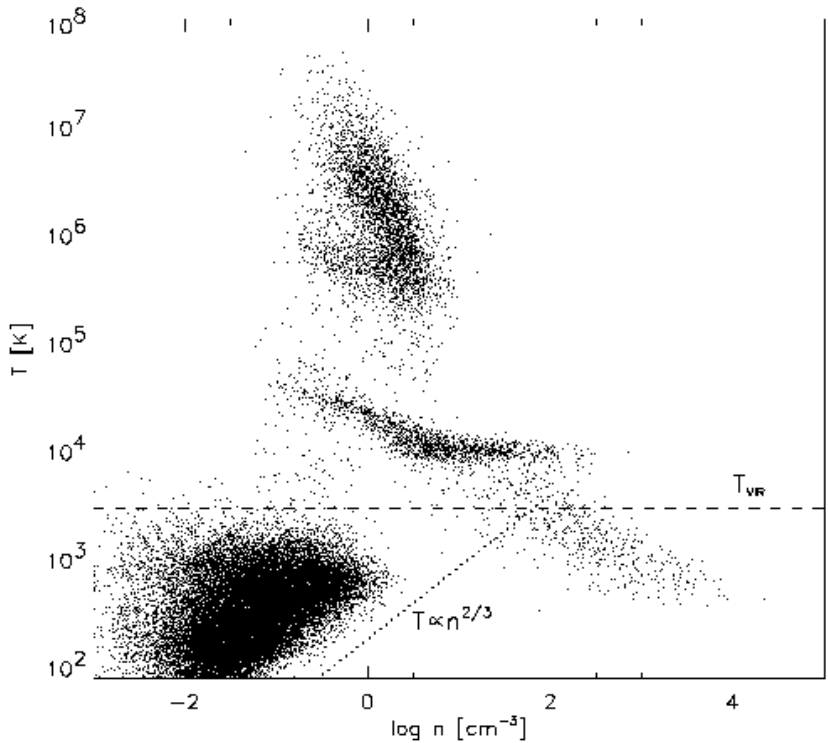,width=10.92cm,height=9.828cm}}
\caption{
Thermodynamic properties of the shocked gas $\sim 10^5$~yr after the explosion.
Gas temperature (in K) vs. log $n$ (in cm$^{-3}$).
{\it Dashed line:} The virial temperature of the minihalo. It is evident that
$T_{\rm vir}\sim 3000$~K is much lower than the gas temperature in the minihalo.
Most of the gas will therefore be expelled from the halo on a 
dynamical timescale.
{\it Dotted line:} Adiabatic behavior. The gas in the outer reaches of the
minihalo, as well as in the general IGM, approximately follows this scaling.
{\it Grey symbols:} Location of the gas that makes up the original
stellar ejecta. This gas is hot and diffuse, can consequently cool
only slowly, and therefore provides the pressure source that drives
the expansion of the bubble.
Notice that a fraction of the gas has been able to cool to $\sim 200$~K
due to the action of molecular hydrogen. (From Bromm et al. 2003.)}
\label{fig1sn}
\end{figure}

\begin{figure} [t]
\centerline{\psfig{file=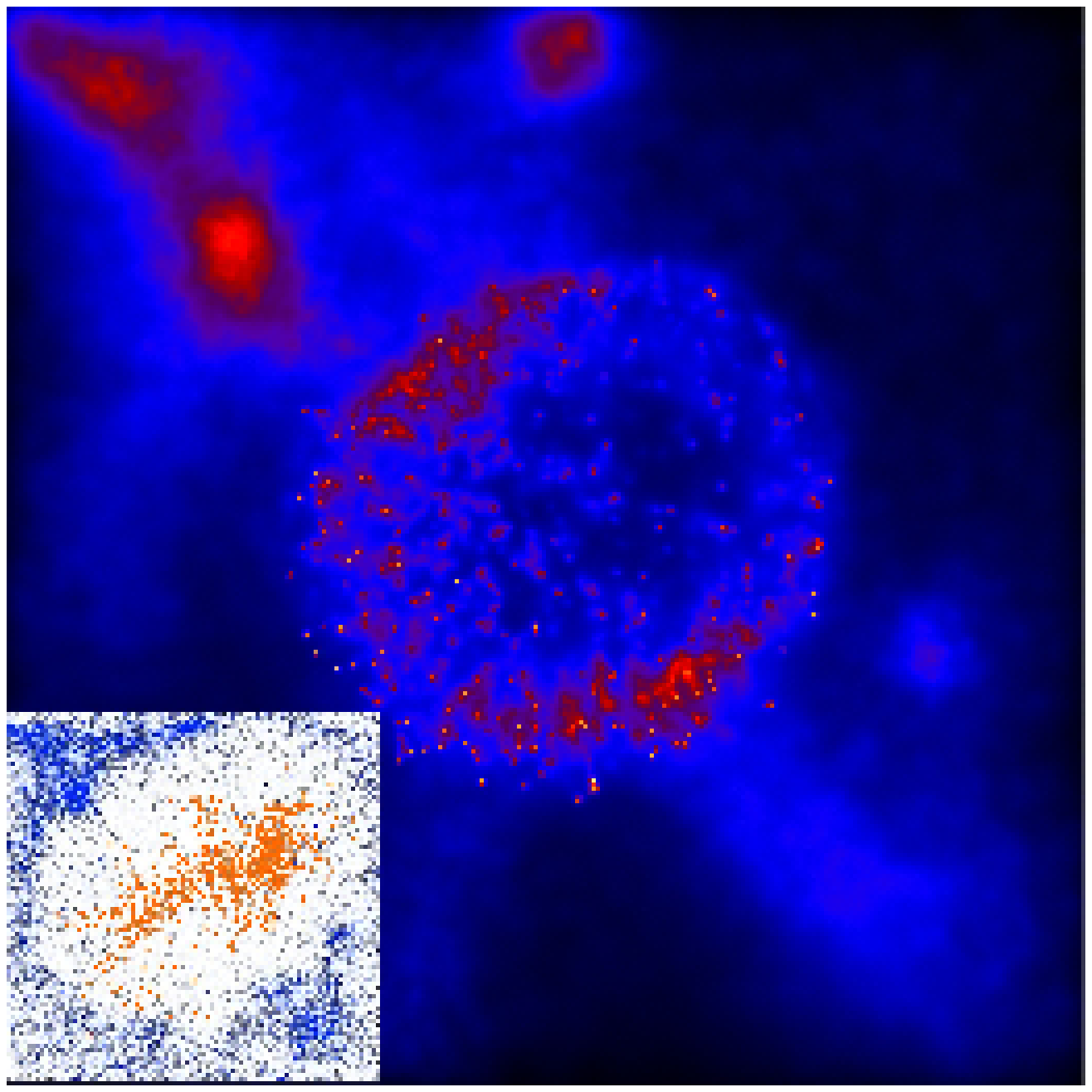,width=10.92cm,height=10.61cm}}
\caption{
Situation $\sim 10^6$~yr after the explosion for the case with
$E_{\rm SN}=10^{53}$ergs. Shown is the projected gas density 
within a box of linear physical size 1 kpc.
The SN bubble has expanded to a radius of $\sim 200$~pc, having evacuated
most of the gas in the minihalo (with a virial radius of $\sim 150$~pc).
The dense shell has fragmented into numerous cloudlets.
{\it Inset:} Metal distribution after 3~Myr. The stellar ejecta ({\it red dots})
trace the metals, and are embedded in pristine, un-enriched gas ({\it blue dots}).
A large fraction ($\ga 90$\%) of the heavy elements has escaped the
minihalo, and filled a significant fraction of the hot bubble.
(From Bromm et al. 2003.)}
\label{fig2sn}
\end{figure}

Recently, Bromm et al. (2003) have presented numerical simulations of the
first supernova explosions at high redshifts ($z\sim 20$). In contrast
to earlier work, both analytical 
(e.g., Larson 1974, Dekel \& Silk 1986, Scannapieco et al. 2002,
Furlanetto \& Loeb 2003)
and numerical (e.g., Mori et al. 2002, Thacker et al. 2002,
Wada \& Venkatesan 2003), their focus is on the minihalos that are the
sites for the formation of the very first stars. These halos,
with masses of $M\sim 10^5 - 10^6 M_{\sun}$ and virializing at $z\ga 20$,
are the sites for the formation of the truly first stars.
It is not obvious, however, whether star formation in more massive
systems would not already be significantly altered by the activity
of previous generations of stars. In particular, PISN explosions
may only have occurred in the minihalos mentioned above.

Bromm et al. (2003) 
assume that there is one single 
Population~III star in the center of the minihalo, and that this star is 
massive enough to explode as a PISN. To span
the possible energy range, explosion energies of
$E_{\rm SN}=10^{51}$~and $10^{53}$ergs are considered, corresponding to stars with masses
$M_{\ast}\simeq 140$ and $260 M_{\sun}$ (Fryer et al. 2001).
The
latter case  marks
the upper mass limit of the PISN regime. 
The explosion energy is inserted as thermal energy, 
such that the explosion is initialized
with conditions appropriate for the adiabatic Sedov-Taylor (ST) phase.
Since the PISN is completely disrupted without leaving a remnant behind, it
is possible to
trace the subsequent fate of the metals by using the SPH particles
that represent the stellar ejecta as markers.

Initially, the blast wave evolves into a roughly uniform medium at radii
$r\la R_{\rm core}\sim 20$~pc, resulting in the usual ST scaling:
$R_{\rm sh}\propto t^{0.4}$. Once the blast wave reaches beyond the core, however,
it encounters an isothermal density profile in the remainder of the halo,
leading to the scaling (e.g., Ostriker \& McKee 1988):
$R_{\rm sh}\simeq 30{\rm \,pc} (E_{\rm SN}/10^{53}{\rm ergs})^{1/3}
(t/10^{5}{\rm yr})^{2/3}$. A few $10^{6}$yr after the explosion, radiative
(inverse Compton) losses become important, and the SN remnant enters its final,
snowplow phase. 
The significance of inverse Compton cooling distinguishes very high-$z$
SN explosions from those occurring in the present-day universe.

After approximately $10^{5}$yr, when $t_{\rm cool} \la R_{\rm sh}/u_{\rm sh}$,
a dense shell begins to form at a radius $\sim 50$~pc. The thermodynamic
properties of the dense shell, the interior bubble, and the surrounding medium
are summarized in Figure~6. The shocked, swept-up gas exhibits three distinct
thermal phases, as can clearly be discerned
in Fig.~6: the first at $T\ga 10^{5}$~K,
the second at $T\sim 10^{4}$~K, and the third at $T\la 10^{3}$~K, with the latter
two corresponding to the dense shell. The cooling from very high temperatures 
to $\sim 10^{4}$~K, and subsequently to $10^{2}$~K, proceeds in non-equilibrium,
such that a fraction of free electrons persists below $\sim 8,000$~K, allowing
the re-formation of H$_{2}$ (e.g., Mac Low \& Shull 1986, Shapiro \& Kang 1987,
Oh \& Haiman 2002). The molecular fraction
in the densest regions asymptotically assumes a `plateau' value
of $f_{{\rm H}_{2}}\sim 3\times 10^{-3}$. Inside the expanding SN remnant
lies a hot pressurized bubble (the grey symbols in Fig.~6) that corresponds
to the original stellar ejecta. This bubble drives the outward motion, and its gas
is slowly cooled by adiabatic expansion and inverse Compton losses.

In Figure~7, the projected gas density is shown $\sim 10^{6}$yr after the
explosion for explosion energy $E_{\rm SN}=10^{53}$ergs. 
The blast wave has succeeded in completely disrupting the
minihalo, dispersing most of the halo gas into the IGM. Comparing $E_{\rm SN}$
with the binding energy of the minihalo, $E_{\rm grav}\sim 10^{51}$ergs,
such an outcome is clearly expected. It depends, however, on the inefficient
radiative processes that cool the hot bubble gas in the initial stages of
expansion. 

The second prominent feature in Fig.~7 is the vigorous fragmentation of
the dense shell into a large number of small cloudlets. 
The emergence of these cloudlets, corresponding to gas at $T\la
10^{3}$K, is not driven by gravity (e.g., Elmegreen 1994 and references
therein), but is instead due to a thermal instability triggered by the
onset of atomic cooling. 
The explosions simulated by Bromm et al. (2003) do not result in the formation of lower-mass
stars, or Population~II.5 stars in the terminology of Mackey et al. (2003).
This is consistent with the prediction
in Mackey et al. (2003) that Population~II.5 stars can only form in dark matter halos
massive enough to be able to cool via atomic hydrogen (see also Salvaterra et al.
2003).
To further test the possibility of SN triggered star formation in the high-redshift
universe, it will be necessary to
explore a range of cosmological
environments, corresponding to different halo masses and collapse redshifts.

What is the fate of the metals that are produced in the PISN progenitor?
PISNe are predicted to have substantial metal yields, of the order
of $y=M_{Z}/M_{\ast}\sim 0.5$, and for the largest stellar masses in
the PISN range, most of $M_{Z}$ is made up of iron (Heger
\& Woosley 2002). Bromm et al. (2003) find that for $E_{\rm SN}=10^{53}$ergs,
$\sim 90$\% of the metals have escaped from the DM minihalo
$4\times 10^{6}$yr after the explosion. The metals have filled most
of the interior hot bubble (see inset of Fig.~7), and there has not yet been
sufficient time for them to significantly mix into the
dense surrounding shell. 
A rough estimate of the resulting metallicity in the
surrounding, contaminated IGM can be obtained as follows: $Z=M_{Z}/M_{\rm gas}
\sim 100 M_{\sun}/10^{5}M_{\sun}\ga 10^{-2}Z_{\sun}$. This value is well above
the critical metallicity threshold, $Z_{\rm crit}\sim 10^{-3.5}Z_{\sun}$, for
enabling the formation of lower-mass stars (see Section~3.5).
The conclusion is
that Population~III stars can continue to form in the pristine gas within the
neighboring halos, as the metals that are dispersed in the explosion 
do not reach them. 

Whereas a single SN explosion might be able to
evacuate most of the gas from a minihalo, the situation 
could be quite different for multiple SN explosions in early dwarf
galaxies of mass $\sim 10^{8} M_{\odot}$ (Mori et al. 2002, Wada \&
Venkatesan 2003). In this case, the simulations indicate that
off-nuclear SN explosions drive inward-propagating blast waves that
act to assemble gas in the central regions of the host galaxy.
The gas in these galaxies may thus be able to survive multiple
SN explosions, and star formation would not be significantly
suppressed due to the mechanical SN feedback. In summary, it appears
that the effect of SN events on star formation in early
galaxies, and on the chemical and thermal evolution of the IGM
depends sensitively on the specific environment.

For the PISN explosion studied by Bromm et al. (2003),
the extent of the metal enriched region, with a physical size of $\sim 1$~kpc,
is comparable to the radius of the relic H\,II region around the minihalo.
In a related paper (Yoshida et al. 2004), it is argued
that if Population~III stars have led to an early partial reionization of the universe,
as may be required by the recent {\it WMAP} results (see Section~5.1),
this will have resulted in a nearly uniform enrichment of the universe to
a level $Z_{\rm min}\ga 10^{-4}Z_{\sun}$ already at $z\ga 15$.
It would indeed be a remarkable feature of the universe, if the first stars
had endowed the IGM with such a widespread, near-universal
level of pre-enrichment (see also Oh et al. 2001). An interesting implication
could be the observability of 
unsaturated metal absorption lines at $z\ga 6$, which
would probe the neutral fraction of the IGM during the epoch of reionization
(Oh et al. 2002).

We end this section by emphasizing that we are only beginning to
understand the complex processes that distributed the first
heavy elements into the IGM, and more sophisticated numerical simulations
are required for further progress.

\section{OBSERVATIONAL SIGNATURE}

An increasing variety of observational probes of the first stars
has been suggested (e.g., Haiman \& Loeb 1997, Oh et al. 2003,
Stiavelli et al. 2003).
Below, we focus on three of them, and here only briefly mention
some of the others (see Barkana \& Loeb 2001 for a comprehensive
discussion). Intriguing possibilities would arise if the Population~III
IMF were sufficiently top-heavy to enable the formation of 
massive black holes (e.g., Madau \& Rees 2001, Islam et al. 2003,
Volonteri et al. 2003). 
Possible consequences might include backgrounds of
gravitational waves (Schneider et al. 2000, de Araujo et al. 2002)
and diffuse neutrinos (Schneider et al. 2002b).
In the latter case, the high-energy (TeV) neutrinos might originate 
in the relativistic jets that are associated with gamma-ray bursts (GRBs),
which are believed to accompany the birth of a black hole (see 
Section 5.3).

The integrated light emitted during the lifetime of Population~III stars
is predicted (Bond et al. 1986, Santos et al. 2002, Salvaterra \& Ferrara 2003)
to significantly contribute to the locally observed cosmic near-infrared
background (e.g., Hauser \& Dwek 2001). A related signature
is the Population~III contribution to the background anisotropy
(Magliocchetti et al. 2003, Cooray et al. 2003). Finally, 21~cm radiation
from neutral hydrogen, both in emission and absorption, could provide 
a unique probe of the conditions in the early minihalos, thus testing 
our assumptions on the initial conditions for Population~III star formation
(e.g., Madau et al. 1997, Tozzi et al. 2000, Carilli et al. 2002, 
Furlanetto \& Loeb 2002, Iliev et al. 2002, 2003, Ciardi \& Madau 2003).
In practice, these observations may be extremely challenging, however,
due to the strong foreground contamination (Di Matteo et al. 2002).

In the following, we discuss three empirical probes of the first stars
in somewhat greater detail, as they have already provided us with
constraints on the nature of the first stars, or soon promise to do so.

\subsection{Reionization Signature from the First Stars}

\begin{figure}[t]
\centerline{\psfig{figure=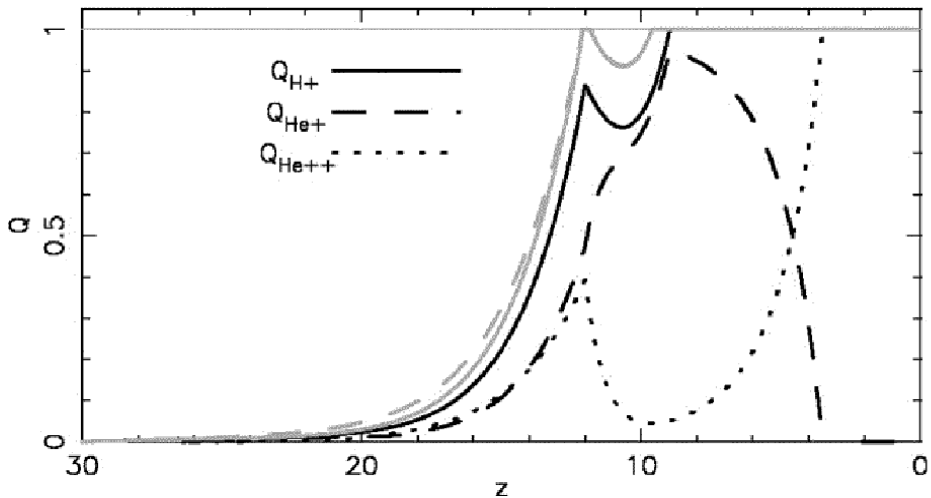,width=5in}}
\caption{Sample history of reionization. Shown are the volume
filling factors of ionized regions ($Q$) as a function of redshift.
For hydrogen ({\it solid line}), a complex evolution is evident, with
an early (partial) reionization at $z\sim 15$ due to massive Population~III
stars. This mode of star formation is self-terminating, such that the
second (complete) reionization at $z\sim 9$ is due to normal
Population~II stars. Also shown is the hypothetical case where
He ionization is ignored ({\it light lines}), as well as the predicted
histories for He$^{+}$ and He$^{++}$ reionization.
(Adapted from Wyithe \& Loeb 2003b.)}
\label{fig9}
\end{figure}
The {\it WMAP} satellite has recently measured, from the CMB polarization
anisotropies (Kaplinghat et al. 2003), the total optical depth
to Thomson scattering: $\tau_e=0.17\pm 0.04$ (Kogut et al. 2003,
Spergel et al. 2003). Such a high value is surprising. We know that the
IGM was completely ionized again at redshifts $z\la 6$ from the
absence of Gunn-Peterson troughs in high-redshift 
Sloan Digital Sky Survey (SDSS) quasars (e.g., Fan
et al. 2003). Scattering from free electrons between $z\sim 0 - 6$
only contributes $\tau_e\simeq 0.05$. The observed excess must
have arisen from the scattering of CMB photons off free electrons
ionized by star formation at higher redshifts. It has been shown
that normal stellar populations (characterized by a standard IMF, or at most
a modestly top-heavy one with characteristic mass $M_c\sim 5 M_{\odot}$)
could have successfully reproduced the measured high optical depth,
if reionization had occurred already at $z\sim 15$ (Ciardi et al. 2003).

Such a solution, however, runs into difficulties when two other
empirical constraints are taken into account. The first one considers
the thermal history of the post-reionization IGM (e.g., Theuns et al. 2002,
Hui \& Haiman 2003). After reionization is completed, the IGM continues
to be heated by photoionization. This heating, however, cannot
compete with the cooling due to adiabatic (Hubble) expansion. In consequence,
the earlier reionization occurred, the lower the IGM temperature at
a given fixed redshift. Observations at $z\sim 3$ indicate that
the last complete overlap of ionizing bubbles (i.e., the last
time the IGM was completely ionized) could not have occurred before
a redshift $z\sim 9$. The second constraint that is difficult to
accommodate within the model of Ciardi et al. (2003), is the detection
of select SDSS quasars at $z\ga 6$ with Gunn-Peterson absorption
in their vicinity. It is not easy to see how islands of neutral gas
could have persisted down to $z\sim 6$, if the IGM had already been
completely reionized at $z\sim 15$. 

To simultaneously account for all the available observational 
constraints thus poses a non-trivial challenge to theory.
It has been suggested that an early generation of very massive
Population~III stars, as predicted by numerical simulations (see Section~3),
could naturally allow for a solution (Cen 2003a,b, Haiman \& Holder 2003,
Somerville \& Livio 2003, Wyithe \& Loeb 2003a,b). 
The high efficiency of producing ionizing
photons per stellar baryon for very massive Population~III stars
(e.g., Tumlinson \& Shull 2000, Bromm et al. 2001b, Schaerer 2002, 2003),
can lead to an early, first episode of at least partial reionization
at $z\sim 15$, thus ensuring a sufficiently large optical depth to
Thomson scattering. As we have discussed in Section~4, Population~III star
formation was self-limiting, both due to the effect of radiation and
the dispersal of heavy elements. After the era of Population~III star
formation at $z\ga 15$, the IGM was able to recombine again, due
to the temporary lull in the production of ionizing photons.
Eventually, normal (Population~II) stars form in more massive
(dwarf-sized) systems, and lead to the final, complete reionization
of the universe at $z\sim 7$. In this case, there is no conflict with the
empirical constraints outlined above. In Figure~8, a sample reionization
history is shown that is representative for the `double-reionization'
models (from Wyithe \& Loeb 2003b). At present, the error bars in
the {\it WMAP} determination of $\tau_e$ are still too large to firmly
establish the need for very massive Population~III stars to have
formed in the high-redshift universe. At the end of its mission, however,
the error on $\tau_e$ is estimated to have gone down to $\pm 0.01$.
If the high current value were to persist, the evidence for a top-heavy
IMF in primordial star formation would have been considerably strengthened.

Cosmological models with reduced small-scale power (see Section~2)
cannot easily be reconciled with a high value of $\tau_e$
(Somerville et al. 2003, Yoshida et al. 2003b,c). To compensate
for the missing contribution from minihalos, one would have to
invoke an extremely high ionizing photon production rate in galaxies
at $z\ga 6$. This could in principle be achieved by maintaining 
a top-heavy IMF over a long period, $6\la z \la 20$. The problem with
such a scenario is the severe overproduction of metals. These would
have to be `hidden' to an improbable degree of precision
so that the formation of massive Population~III stars 
were not terminated too early on (Yoshida et al. 2004). 
The study of the first stars thus has the exciting potential
to provide astrophysical constraints on the nature of dark matter
particles.

\subsection{Stellar Archaeology: Relics from the End of the Dark Ages}

It has long been realized that the most metal-poor stars found in
our cosmic neighborhood would encode the signature from the first stars
within their elemental abundance pattern (e.g., Bond 1981, Beers et al. 1992).
For many decades, however, the observational search has failed
to discover a truly first-generation star with zero metallicity. Indeed,
there seemed to have been an observational lower limit of [Fe/H] $\sim -4$ 
(e.g., Carr 1987, Oey 2003).
In view of the recent theoretical prediction that most Population~III stars
were very massive, with associated lifetimes of $\sim 10^6$~yr, the failure
to find any `living' Population~III star in the Galaxy is not surprising,
as they all would have died a long time ago (e.g., Hernandez \& Ferrara 2001).
Furthermore, theory has
predicted that star formation out of extremely low-metallicity gas,
with $Z\la Z_{\rm crit}\sim 10^{-3.5}Z_{\odot}$ (see Section 3.5), would
be essentially equivalent to that out of truly primordial gas.
Again, this theoretical prediction was in accordance with the apparent
observed lower-metallicity cutoff.

\begin{figure}[t]
\centerline{\psfig{figure=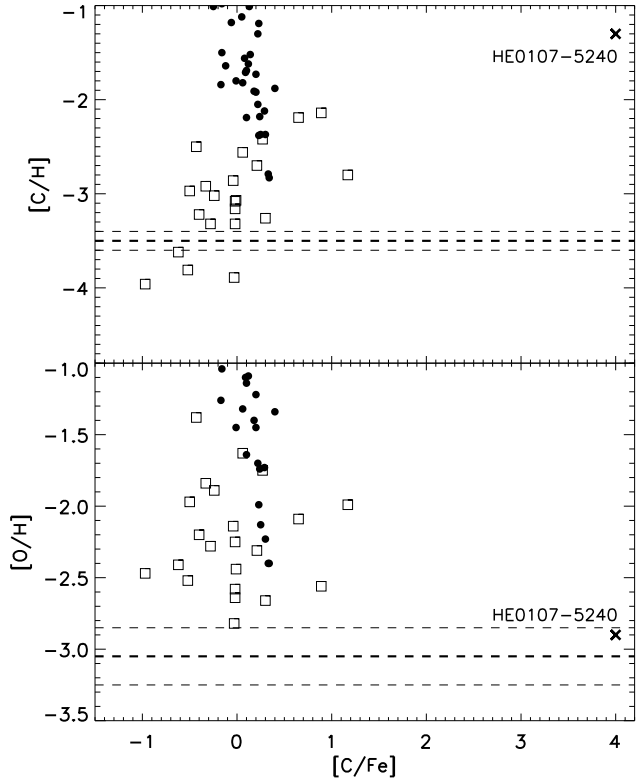,height=4.5in}}
\caption{Observed abundances in low-metallicity Galactic halo stars.
For both carbon ({\it upper panel}) and oxygen ({\it lower panel}),
filled circles correspond to samples of dwarf and subgiant stars
(from Akerman et al. 2003), and open squares to a sample of
giant stars (from Cayrel et al. 2003). The dashed lines indicate
the predicted critical carbon and oxygen abundances (see Section 3.5).
Highlighted ({\it cross}) is the location
of the extremely iron-poor  giant star HE0107-5240.
(Adapted from Bromm \& Loeb 2003c.)}
\label{fig10}
\end{figure}

Recently, however, this simple picture has been challenged by the
discovery of the star HE0107-5240 with a mass of $0.8 M_{\odot}$
and an {\it iron} abundance of ${\rm [Fe/H]} = -5.3$ 
(Christlieb et al. 2002). This finding indicates that at least
some low mass stars could have formed out of extremely low-metallicity gas.
Does the existence of this star invalidate the theory of a metallicity
threshold for enabling low-mass star formation? As pointed out 
by Umeda \& Nomoto (2003), a possible explanation could lie in
the unusually high abundances of carbon and oxygen in HE0107-5240.
As discussed in Section~3.5, Bromm \& Loeb (2003c) have demonstrated
that indeed C and O may have been the main drivers for the Population~III to II
transition.

In Figure~9 we compare the theoretical thresholds derived by Bromm \& Loeb
(2003c) to the observed C and O
abundances in metal-poor dwarf (Akerman et al. 2003) 
and giant (Cayrel et al. 2003) stars in
the halo of our Galaxy.  As can be seen, all data points lie above the
critical O abundance but a few cases lie below the critical C threshold.
All of these low mass stars are consistent with the model since the
corresponding O abundances lie above the predicted threshold.  The
sub-critical [C/H] abundances could have either originated in the
progenitor cloud or from the mixing of CNO-processed material (with
carbon converted into nitrogen) into the stellar atmosphere during the red
giant phase. To guard against this a~posteriori processing of C (and to a
lesser degree of O), dwarf stars should be used as they provide a more
reliable record of the primordial abundance pattern that was present at
birth. The current data on dwarf stars (filled symbols in Fig. 9) does not
yet reach C and O abundances that are sufficiently low to probe the
theoretical cooling thresholds.  Note also that the extremely iron-poor
star HE0107-5240 has C and O abundances that both lie above the respective
critical levels. The formation of this low mass star ($\sim 0.8 M_{\odot}$)
is therefore consistent with the theoretical framework considered
by Bromm \& Loeb (2003c).

The lessons from stellar archaeology on the nature of the first stars
are likely to increase in importance, since greatly improved, large
surveys of metal-poor Galactic halo stars are under way, or are currently
being planned.

\subsection{Gamma-Ray Bursts as Probes of the First Stars}

GRBs are the brightest electromagnetic explosions
in the universe, and they should be detectable out to very high redshifts
(e.g., Wijers et al. 1998, Blain \& Natarajan 2000,
Ciardi \& Loeb 2000, Lamb \& Reichart 2000, Bromm \& Loeb 2002,
Choudhury \& Srianand 2002).
Although the nature of the central engine that
powers the relativistic jets is still debated, recent evidence
indicates that GRBs trace the formation of massive stars
(e.g., Bloom et al. 2002, and references therein).
Since the first stars are
predicted to be predominantly very massive, their death might possibly
give rise to GRBs at very high redshifts (e.g., Schneider et al. 2002b). 
A detection of the
highest-redshift GRBs would probe the earliest epochs of star
formation, one massive star at a time 
(e.g., Barkana \& Loeb 2003, M\'{e}sz\'{a}ros \& Rees 2003). 
The upcoming {\it Swift}
satellite\footnote{See http://swift.gsfc.nasa.gov.}, planned for
launch in 2004, is expected to detect about a hundred GRBs per
year. The redshifts of high-$z$ GRBs can be easily measured through
infrared photometry, based on the Gunn-Peterson trough in their spectra due to
Ly$\alpha$ absorption by neutral intergalactic hydrogen along the line of sight.
{\it Which fraction of the detected bursts will originate at
redshifts $z\ga 5$?}

\begin{figure}[t]
\centerline{\psfig{figure=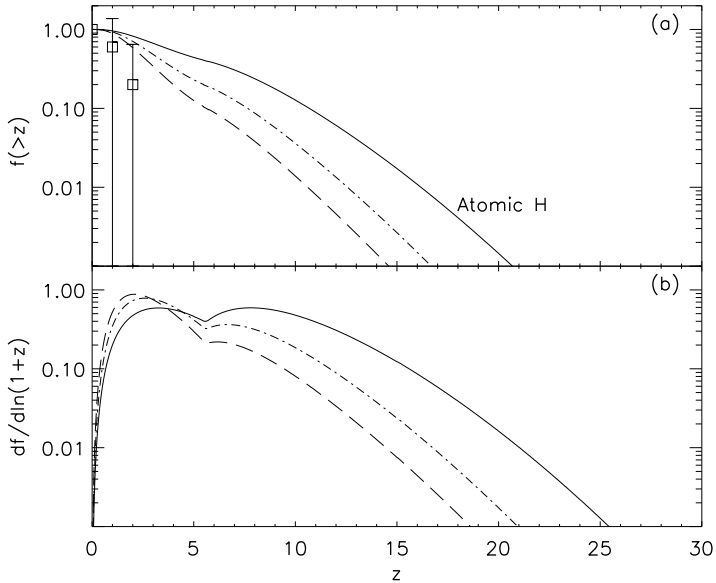,width=4.in}}
\caption{Redshift distribution of all GRBs in comparison with
measurements in flux-limited surveys.  ({\it a})
Fraction of bursts that originate at a redshift higher than $z$
vs. $z$.  The data points reflect $\sim 20$ observed redshifts to
date.  ({\it b}) Fraction of bursts per logarithmic interval of
$(1+z)$ vs. $z$.  {\it Solid lines:} All GRBs for star formation in
halos massive enough to allow cooling via lines of atomic hydrogen.
The calculation assumes that the GRB rate is proportional (with a
constant factor) to the star formation rate at all redshifts.  {\it
Dot-dashed lines:} Expected distribution for {\it Swift}.  {\it
Long-dashed lines:} Expected distribution for BATSE.  The curves for
the two flux-limited surveys are rather uncertain because of the
poorly-determined GRB luminosity function.  
(From Bromm \& Loeb 2002.)}
\end{figure}

To assess the utility of GRBs as probes of the first stars, one has to 
calculate the expected redshift distribution of GRBs 
(e.g., Bromm \& Loeb 2002, Choudhury \& Srianand 2002).
Under the assumption that the GRB rate is simply proportional to the
star formation rate, Bromm \& Loeb (2002) find
that about a quarter of all GRBs detected
by {\it Swift}, will originate from a redshift $z\ga 5$ (see
Fig.~10). This estimate takes into account the detector sensitivity, but it
is still very uncertain because of the poorly
determined GRB luminosity function.  The rate of
high-redshift GRBs may, however, be significantly suppressed if the early
massive stars fail to launch a relativistic outflow. This is
conceivable, as metal-free stars may experience negligible mass loss
before exploding as a supernova (Baraffe et al. 2001). 
They would then retain their massive
hydrogen envelope, and any relativistic jet might be quenched before
escaping from the star (Heger et al. 2003).

If high-redshift GRBs exist, the launch of {\it Swift} 
will open up an exciting new window into the cosmic dark
ages.  Different from quasars or galaxies that fade with
increasing redshift, GRB afterglows maintain a roughly constant
observed flux at different redshifts for a fixed observed time lag
after the $\gamma$-ray trigger (Ciardi \& Loeb 2000). The increase in the
luminosity distance at higher redshifts is compensated by the fact
that a fixed observed time lag corresponds to an intrinsic time
shorter by a factor of $(1+z)$ in the source rest-frame, during which
the GRB afterglow emission is brighter. This quality makes GRB
afterglows the best probes of the metallicity and ionization state of
the intervening IGM during the epoch of reionization. In contrast
to quasars, the UV emission from GRBs has a negligible effect on the
surrounding IGM (since $\sim 10^{51}$ ergs can only ionize $\sim 4\times
10^4M_\odot$ of hydrogen). Moreover, the host galaxies of GRBs induce a much
weaker perturbation to the Hubble flow in the surrounding IGM, compared to
the massive hosts of the brightest quasars (Barkana \& Loeb 2003).
Hence, GRB afterglows offer the ideal probe (much better than quasars or
bright galaxies) of the damping wing of the Gunn-Peterson trough
(Miralda-Escud\'{e} 1998)
that signals the neutral fraction of the IGM as a function
of redshift during the epoch of reionization.

\section{EPILOGUE: THE ROAD AHEAD}

One of the fascinating aspects of the recent research on the first stars
has been the confluence of many hitherto unrelated subfields.
To fully explore its rich potential, we have to utilize
both in situ, high-redshift studies, as well as near-field 
cosmological data (Freeman \& Bland-Hawthorn 2002). The latter approach refers
to the stellar archaeology of observing the chemical abundance patterns
in extremely metal-poor Galactic halo stars (see Section 5.2).
Furthermore, the study of the first stars lies at the intersection
of star formation theory with cosmological structure formation, 
where the scales for the phenomena overlap. Similarly, there is also
an overlap between star and galaxy formation, because at this early
time the scale of galaxies and the scale of star-forming clouds
had not yet become clearly separated. Therefore, we can investigate
all of these phenomena in their most elementary form, before
their scales had become widely separated and before the universe
had evolved into the vastly more complicated structure that we observe
today, where there is a huge range of scales at play.

Most of the work so far has been theoretical, but the field is driven
by the prospect of upcoming observational probes that will soon test
the theoretical predictions. The {\it WMAP} satellite has already provided
important clues, and {\it Swift} will soon do so as well. Most importantly,
NASA's planned successor mission to the Hubble Space Telescope, the {\it JWST},
to be launched around 2011, is a prime driver behind current cosmological
research. The {\it JWST} is designed to image the first stars, supernovae, and
quasars. In thinking ahead, pondering about the first glimpses that
we have gained into the cosmic dark ages, one wonders what exciting 
discoveries are awaiting us.

\vspace{.4in}
\noindent
{\bf ACKNOWLEDGMENTS}
\vspace{.1in}

\noindent
We express our gratitude to Paolo Coppi
for the many discussions on
the work discussed in this review.
VB thanks Avi Loeb and Lars Hernquist 
at the Harvard-Smithsonian Center for Astrophysics
for support from NSF grant AST 00-71019, as well as
Cathie Clarke at the Institute of Astronomy in Cambridge
from the European Community's Research Training Network
under contract HPRN-CT-2000-0155, `Young Stellar Clusters'.

\vspace{.5in}


\end{document}